\documentclass[aps, prd, reprint, groupedaddress, nofootinbib]{revtex4-1}

\usepackage{aas_macros}
\usepackage{bm}
\usepackage{times}
\usepackage{graphicx}
\usepackage{color}
\usepackage{amsmath}
\usepackage{amssymb}
\usepackage{fancyvrb}
\usepackage{url}
\usepackage{verbatim}
\usepackage[utf8]{inputenc}
\usepackage[varg]{txfonts}

\renewcommand{\cite}{\citep}

\newcommand{\tcov}{\mat{S}_\mathrm{T}}
\newcommand{\scov}{\mat{S}_\mathrm{S}}
\DeclareMathOperator{\Tr}{Tr\,}
\newcommand{\nside}{N_\mathrm{side}}
\newcommand{\ukarcmin}{\mu{\rm K}\,{\rm arcmin}}
\newcommand{\vect}{\boldsymbol}
\newcommand{\mat}{\boldsymbol}
\newcommand{\bmode}{$B$-mode}
\newcommand{\bmodes}{$B$-modes}

\newcommand{\qed}{\nobreak \ifvmode \relax \else
      \ifdim\lastskip<1.5em \hskip-\lastskip
      \hskip1.5em plus0em minus0.5em \fi \nobreak
      \vrule height0.75em width0.5em depth0.25em\fi}

\begin{document}
\VerbatimFootnotes

\author{Eric R. Switzer}
\affiliation{NASA Goddard Space Flight Center, Greenbelt, Maryland 20771, USA}
\author{Duncan J. Watts}
\affiliation{Department of Physics and Astronomy, The Johns Hopkins University, 3400 North Charles Street, Baltimore, Maryland 21218, USA}

\title{Robust likelihoods for inflationary gravitational waves from maps of cosmic microwave background polarization}

\begin{abstract}
The \bmode\ polarization of the cosmic microwave background provides a unique window into tensor perturbations from inflationary gravitational waves. Survey effects complicate the estimation and description of the power spectrum on the largest angular scales. The pixel-space likelihood yields parameter distributions without the power spectrum as an intermediate step, but it does not have the large suite of tests available to power spectral methods. Searches for primordial \bmodes\ must rigorously reject and rule out contamination. Many forms of contamination vary or are uncorrelated across epochs, frequencies, surveys, or other data treatment subsets. The cross power and the power spectrum of the difference of subset maps provide approaches to reject and isolate excess variance. We develop an analogous joint pixel-space likelihood. Contamination not modeled in the likelihood produces parameter-dependent bias and complicates the interpretation of the difference map. We describe a null test that consistently weights the difference map. Excess variance should either be explicitly modeled in the covariance or be removed through reprocessing the data.
\end{abstract}

\maketitle

\section{Introduction}

Systematic error control and rejection are central considerations of cosmic microwave background (CMB) instrument design and data analysis. The reionization feature of inflationary gravitational waves \citep{1997PhRvD..55.7368K, 1997PhRvL..78.2054S, 1997NewA....2..323H} and improved constraints on the optical depth $\tau$ \citep{2015PhRvD..92l3535A} are of great interest, and require analysis of CMB polarization on the widest angular scales of the sky. These measurements are susceptible to contamination because they relate instrument response and foregrounds across the largest angular and temporal separations. 

A well-established approach to detect or reject systematics splits the data into epochs, frequencies, surveys or other subsets across which a contaminant varies or is uncorrelated. This general approach can take the form of a cross-spectral estimator \citep{2005MNRAS.358..833T} or power spectrum null tests across a variety of difference maps. The maps can be split and subtracted to check for particular instrumental effects such as time constants (e.g. \citep{2015ApJ...811..126B}) or to get uncorrelated realizations of detector noise or atmospheric fluctuations (e.g., \citep{2013ApJ...779...86S, 2014JCAP...04..014D}). Contamination that is uncorrelated between the maps does not produce bias in the cross power, but it does boost errors. Here, we consider robustness and bias in the pixel-space likelihood and develop an approach analogous to the cross power or difference map null test.

Anisotropy spectral analysis compresses map information by exploiting the Gaussianity and statistical isotropy of the CMB signal. A survey is limited to fractions of the sky by its scan strategy and galactic contamination. Truncation has two consequences for the power spectrum. The estimate $\hat C_\ell$ is the sum of quadratic products of normally distributed map variations, making $P(C_\ell | \vect{x})$ (given map data vector $\vect{x}$) non-Gaussian unless there are sufficient modes available to be in the central limit. On a partial sky, the spherical harmonics for intensity and polarization are an incomplete basis, which results in correlations between $\ell$ \citep{2014PhRvD..89f3008G, 2002ApJ...567....2H}, and $E$- and \bmode\ polarization mixing \citep{2003PhRvD..67b3501B}. Polarization power spectra through pseudo-$C_\ell$ \citep{1973ApJ...185..413P, 2002ApJ...567....2H} and related quadratic methods have well-established procedures for deriving the $P(C_\ell | \vect{x})$ \citep{2006MNRAS.370..343E, 2009PhRvD..79h3012H, 2009PhRvD..79l3515G}. \citet{2015MNRAS.453.3174M} recently developed an implementation of a complete probability distribution function for the cross power spectrum on large scales, including the effects above.

An alternative approach resolves the challenges of spectral estimation by determining the cosmological parameters directly from the map \citep{2007ApJS..170..335P, 2011ApJ...737...78K, 2013ApJ...771...12F, 2015ApJ...814..103W, 2015arXiv150702704P}. The pixel-space likelihood approach has several advantages. Cuts and variation in coverage on a partial sky are included in the pixel covariance model and do not require simulations or analytic treatment of multipole correlation or polarization mixing. The likelihood represents the complete information contained in maps with Gaussian signal and noise, so it can achieve the lowest-variance estimates of the parameters. It accommodates some classes of foreground subtraction and self-consistently propagates parameter errors \citep{2011ApJ...737...78K, 2015ApJ...814..103W, 2015arXiv150702704P}. The likelihood avoids the need to represent a complete joint (non-Gaussian, correlated) probability distribution of the $C_\ell$'s for $\ell$ on angular scales comparable to the survey size. Finally, the likelihood does not require reference to fiducial model parameters. There are well-established procedures for combining information from pixel-space likelihoods of large scales and spectral analysis of small scales \citep{2007ApJS..170..335P, 2015arXiv150702704P}. 

Drawing an analogy to the power spectrum null test, we can estimate cosmological parameters from the likelihood of a difference between two maps from subsets of the data. The sky signal drops out in the difference, and only noise or variable contamination remains. If there is no contamination, the posterior distribution of cosmological parameters will be consistent with data containing only detector noise. A null test based on the likelihood of the difference may fail to be informative when contamination does not match covariance in the data model. In this case, excess variance produces parameter-dependent bias through the effective weighting of the contaminants in map space. In the difference map, there is no signal, and the contaminant is weighted differently than in the sum.

We develop the joint likelihood analogy to the cross power and null test and propose a reweighting method for difference maps. Reweighting gives a consistent interpretation of bias from contamination in the sum and difference. However, excess variance (misspecification) also produces bias in the width of the posterior parameter distribution. Contamination must be treated either in a reanalysis of the data or as a new term in the likelihood model.

The methods developed here apply to experiments and missions specifically seeking the largest scales on the sky (Advanced ACTPol \citep{2016JLTP..184..772H}, CLASS \citep{2014SPIE.9153E..1IE}, GroundBIRD \citep{2012SPIE.8452E..1MT}, LSPE \citep{2012SPIE.8446E..7AA}, PIPER \citep{2014SPIE.9153E..1LL}, and QUIJOTE \citep{2014arXiv1401.4690L}) and missions (CORE+\,\footnote{\tt http://conservancy.umn.edu/handle/11299/169642}, Inflation Probe \citep{2014arXiv1401.3741K}, LiteBIRD \citep{2014JLTP..176..733M}, and PIXIE \citep{2011JCAP...07..025K}), but also to the largest angular scales of surveys on smaller regions.

Section~\ref{sec:likeadv} reviews the likelihood approach and its relation to spectral methods.
Section~\ref{sec:detnoise} extends the likelihood to include the detector noise amplitude as a nuisance parameter. This setting demonstrates properties of an incomplete covariance model.
Section~\ref{sec:jointlikelihood} reviews the cross-spectral estimator and defines an analogous joint likelihood between data splits.
Section~\ref{ssec:rewighted} describes the joint estimator with unknown contamination covariance and develops a re-weighted difference map null test.
Section~\ref{sec:summary} summarizes the approach.

\section{The likelihood of CMB maps}
\label{sec:likeadv}

The Gaussian log-likelihood $\mathcal L\equiv-\ln P(\vect{x} | \vect{\Theta})$ for map data $\vect{x}$ given parameters $\vect{\Theta}$ is
\begin{eqnarray}
\label{eqn:likeform}
2 \mathcal L &=& \Tr[\ln \mat{C}(\vect{\Theta}) + \mat{C}(\vect{\Theta})^{-1} \mat{D}(\vect{x}, \vect{\Theta})] \\
\mat{D}(\vect{x}, \vect{\Theta}) &\equiv& [\vect{x} - \vect{\mu}(\vect{\Theta})] [\vect{x} - \vect{\mu}(\vect{\Theta})]^T\nonumber,
\end{eqnarray}
where $\mat{C}$ and $\vect{\mu}$ are the covariance and mean of the map vector $\vect{x}$. In subsequent equations, we assume implicit parameter dependence in the mean and covariance model or emphasize dependence on individual model parameters. 

\begin{figure}[htb]
\includegraphics[scale=0.35]{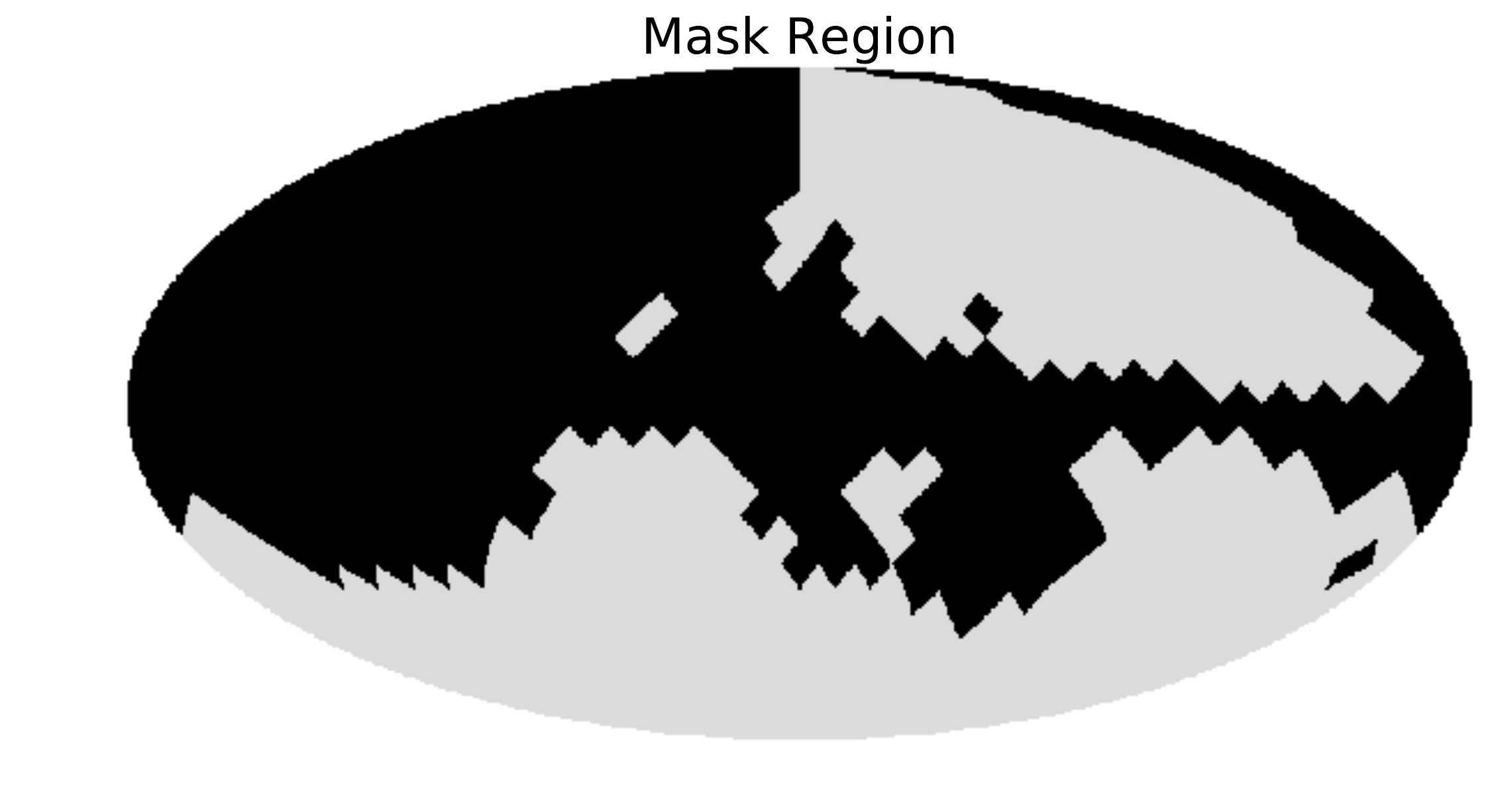}
\caption{Mask used in the simulations here ($\nside=8$, giving $f_{\rm sky} \approx 0.5$), representative of the region accessible from the Atacama, as well as masking the galaxy through WMAP's P06 map.
\label{fig:skymask}}
\end{figure}

The data vector for CMB polarization is $\vect{x}^T \equiv (\vect{x}_Q^T, \vect{x}_U^T)$, a stack of Stokes $Q$ and $U$ maps. Stokes $I$ intensity can also be added for complete $TT$, $TE$, $EE$, and $BB$ two-point information, but models and simulations here use only Stokes $Q$ and $U$ for simplicity.  The maps used in the likelihoods are smoothed at $\theta_{\rm FWHM} = 15^\circ$ and binned onto $\nside=8$ \citep{2005ApJ...622..759G}, encompassing multipoles $2 \leq \ell \leq 23$. The mask region is defined by {\it WMAP} P06 \citep{2007ApJS..170..335P} and a declination limit of $-73^\circ < \delta < 27^\circ$ available to wide-area surveys in the Atacama, such as Advanced ACTPol \citep{2016JLTP..184..772H} and CLASS \citep{2015ApJ...814..103W} in the near term. Combined, $f_{\rm sky} \approx 0.5$, shown in Fig.\,\ref{fig:skymask}. 

The total covariance $\mat{C}(\vect{\Theta})$ is the sum of cosmological signal and noise, $\mat{C}(\vect{\Theta}) = \mat{S}(\vect{\Theta}) + \mat{N}(\vect{\Theta})$. The signal covariance matrix is defined from the power spectrum $C_\ell(\vect{\Theta})$ through \citep{2015arXiv150702704P}
\begin{equation}
\mat{S}(\vect{\Theta}) = \sum_\ell \sum_{XY} C_\ell^{XY}(\vect{\Theta}) \mat{P}_\ell^{XY},
\label{eqn:sigcovmodel}
\end{equation}
where our $XY$ sum extends over only $EE$ and $BB$ to predict the Stokes $Q$ and $U$ covariance. $\mat{P}_\ell^{XY}$ is described in \citet{2001PhRvD..64f3001T} and includes effects of the $\theta_{\rm FWHM} = 15^\circ$ smoothing and incomplete sky coverage. For simplicity, we fix $\Lambda$CDM cosmological parameters \citep{2013ApJS..208...19H} throughout, but these could be jointly estimated in the likelihood. Conclusions are not dependent on the base cosmology at the currently available precision. The input data $\vect{x}$ are generated using \texttt{synfast} \citep{2005ApJ...622..759G} plus a Gaussian random detector noise and are consistent with the covariance in Eq.\,(\ref{eqn:sigcovmodel}). In subsequent sections, we add contamination to the maps to study departures from the model. 

For the purpose of demonstrating the \bmode\ constraint, we use a reference survey with noise RMS amplitude $10\,\ukarcmin$. This noise level is typical of per-band sensitivities of next-generation experiments in their prime CMB science band (see e.g. \citet{2016JCAP...03..052E} and \citet{2016MNRAS.458.2032R} for summaries). Throughout, ``detector noise" refers to the variance in the map attributable to the detectors, which we assume is uncorrelated between map pixels. 

To simulate the distribution of experimental outcomes, we find maximum likelihood (ML) parameters across Monte Carlo realizations of data sets using the limited-memory Broyden-Fletcher-Goldfarb-Shanno algorithm \citep{byrd1995limited,zhu1997algorithm}. 

\subsection{Relation to quadratic methods}
\label{ssec:likequad}

The Newton-Raphson approach provides an analytic expression to iterate to find the maximum likelihood as \citep{1998PhRvD..57.2117B, 2004MNRAS.349..603E} 
\begin{eqnarray}
\label{eq:nriter}
\delta \theta_i &=& \sum_j F_{ij}^{-1} \frac{1}{2} \Tr \left [ (\vect{x} \vect{x}^T -\mat{C}) \left ( \mat{C}^{-1} \mat{C}_{,j} \mat{C}^{-1} \right ) \right ] \\
F_{ij} &\equiv& \frac{1}{2} \Tr(\mat{C}^{-1} \mat{C}_{,i} \mat{C}^{-1} \mat{C}_{,j}),
\end{eqnarray}
where commas denote partial derivatives as $\mat{C}_{,i} \equiv \partial \mat{C} / \partial \theta_i$. This expression gives the update to the parameter $\theta_i$ among the parameters in $\vect{\Theta} = \{ \theta_1, ...\}$. Equation\,(\ref{eq:nriter}) evaluates all covariances and derivatives at the current iteration, and the resulting $\delta \theta_i$ gives the vector of changes $\delta \vect{\Theta}$ to iterate to the maximum likelihood. Each iteration is driven by $\vect{x} \vect{x}^T -\mat{C}(\vect{\Theta})$, the difference between the covariance model at that iteration and the outer product of the data.

The covariance model $\mat{C} = \alpha \mat{C}_{,\alpha} + \mat{N}$ provides a simple example of quadratic estimation of the covariance amplitude $\alpha$. Let the first guess be $\alpha=0$. Then the first step toward the maximum likelihood value of $\alpha$ is
\begin{eqnarray}
\label{eqn:quadest}
\hat \alpha &=& \frac{\vect{x}^T \mat{Q}_{\alpha} \vect{x} - b}{\Tr(\mat{Q}_{\alpha} \mat{C}_{,\alpha})}~~~~b=\Tr(\mat{Q}_{\alpha} \mat{N}) \\
\mat{Q}_{\alpha} &=& \mat{N}^{-1} \mat{C}_{,\alpha} \mat{N}^{-1}.
\end{eqnarray}
The quadratic term $\vect{x}^T \mat{N}^{-1} \mat{C}_{,\alpha} \mat{N}^{-1} \vect{x}$ inverse-noise weights the map data ($\mat{N}^{-1} \vect{x}$) and dots across the covariance structure $\mat{C}_{,\alpha}$. The ``noise bias" $b$ removes the contribution of noise $\mat{N}$. The denominator is a normalization that ensures that the expectation value is
\begin{eqnarray}
\langle \hat \alpha \rangle &=& \frac{\Tr(\alpha \mat{Q}_{\alpha} \mat{C}_{,\alpha} + \mat{Q}_{\alpha}\mat{N}) - \Tr(\mat{Q}_{\alpha} \mat{N})}{\Tr(\mat{Q}_{\alpha} \mat{C}_{,\alpha})} = \alpha.
\end{eqnarray}

The quadratic estimator with full $\mat{C}^{-1}$ weights is the minimum-variance estimate of Gaussian covariance amplitude parameters \citep{1997PhRvD..55.5895T, 2001PhRvD..64f3001T}. Quadratic methods are commonly used to estimate the anisotropy spectrum \citep{2002ApJ...567....2H, 2004MNRAS.348..885E, 2004MNRAS.349..603E, 2006MNRAS.370..343E, 2009MNRAS.400..463G, 2011MNRAS.414..823R, 2015ApJS..221....5G}. In this case, the parameters of the covariance model are the $C_\ell$'s themselves or band powers.

The quadratic approach to the maximum likelihood provides some analytic intuition about the behavior of the maximum likelihood (Sec.\,\ref{ssec:bbias}) and relation of the joint pixel-space likelihood to the cross power (Sec.\,\ref{sec:jointl} and Appendix~\ref{app:jointcross}). 

\subsection{Applicability of the likelihood approach}
\label{ssec:applike}

The pixel-space likelihood is analytically simple at the expense of being computationally intensive and structurally rigid. Evaluation of the likelihood requires the specification and inverse of the $N_{\rm pix} \times N_{\rm pix}$ covariance matrix of the maps, which is numerically expensive. The pixel-space likelihood has therefore seen greatest use in extracting information from the largest scales in the survey, which span $O(1000)$ pixels. Note that the pixel-space analysis is generally useful for modeling signal covariance on angular scales approaching the size of the survey, not just at low-$\ell$.

Given map data and a model, the likelihood is a self-sufficient ``black box" to determine the cosmological parameters. As long as the data model is accurate, the likelihood gives the probability distribution of the parameters. Spectral methods provide greater freedom, such as choosing spatial weightings, removing foregrounds in advance, and throwing out spatial modes. Freedom also carries the responsibility of propagating treatments to the final parameter distribution. Choices in data weighting and filtering other than $\mat{C}^{-1}$ move away from optimality. In contrast, the likelihood has only the fixed model and does not differentiate between the signal structure covariance and data weighting or filtering.

Covariance in the observed data that is neither modeled nor isolated could produce a spurious detection of \bmodes\ or bias the true value. Some contaminant covariance structure may be known accurately in advance. For example, the variation of detector noise across the map can be modeled from the survey coverage and data cuts. However, the amplitude of that noise may not be known from only laboratory measurements or characterization of the time domain data. It is straightforward to include detector noise amplitude in the likelihood's covariance model and to fit it jointly with the cosmological parameters. Fitting for the amplitude removes that source of bias from the cosmological parameters and gives confidence regions that reflect the full covariance.

\section{Bias from excess detector noise}
\label{sec:detnoise}

A simple covariance model for CMB polarization data is the sum of tensor and scalar cosmological contributions and a known noise, or \citep{2011ApJ...737...78K, 2015ApJ...814..103W}
\begin{equation}
\mat{C}(\{r, s\}) = r \tcov + s \scov + \mat{N},
\label{eqn:simplecov}
\end{equation}
where the tensor-to-scalar ratio $r$ multiplies the tensor covariance structure $\tcov$, the scalar amplitude $s\equiv A_s/A_{s,0}$ multiplies scalar covariance $\scov$, and $\mat{N}$ is fixed detector noise. $\tcov$ and $\scov$ are derived from the sum on EE and BB in Eq.\,(\ref{eqn:sigcovmodel}) for fixed $\Lambda$CDM cosmological parameters \citep{2013ApJS..208...19H}.

\begin{figure}[htb]
\includegraphics[scale=0.55]{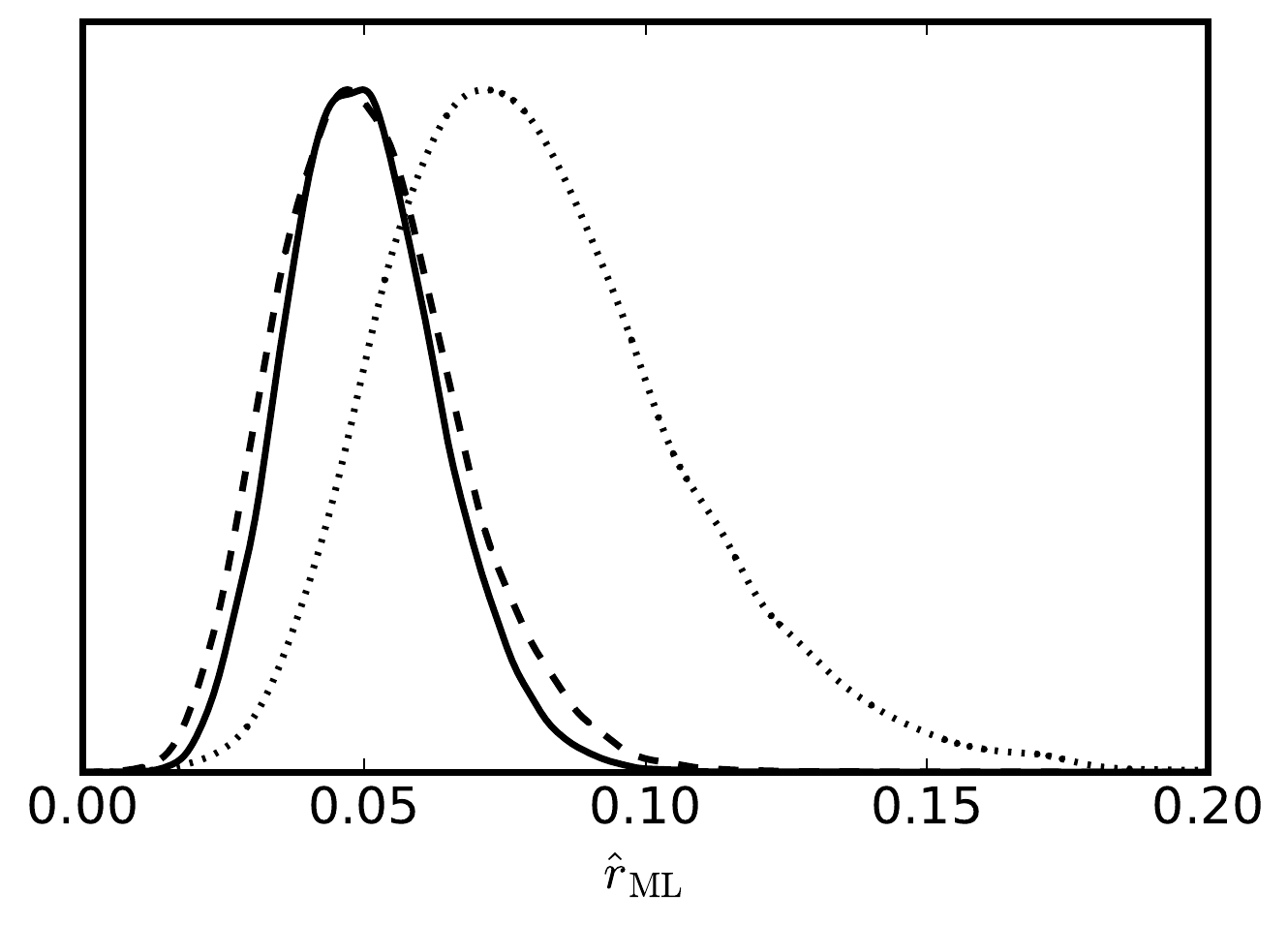}
\caption{The Monte Carlo distribution of the maximum likelihood tensor-to-scalar ratio $r$ for several scenarios of detector noise. {\it Solid curve:} the distribution of $r$ when both the map and the covariance model have $10\,\ukarcmin$ noise, and a true input $r=0.05$. {\it Dotted curve:} the biased distribution of $r$ when the map noise is $13\,\ukarcmin$ while the likelihood's covariance model assumes $10\,\ukarcmin$. {\it Dashed curve:} the distribution of $r$ with a map noise level of $13\,\ukarcmin$, and the likelihood models the detector noise amplitude jointly with the signal. Jointly fitting for the noise recovers the input $r$ value and represents the larger uncertainty in $r$ from $13\,\ukarcmin$ noise vs. the $10\,\ukarcmin$ reference case. 
\label{fig:detnoise_fit}}
\end{figure}

Figure\,\ref{fig:detnoise_fit} shows the Monte Carlo maximum likelihood distribution of $r$ for an input $r=0.05$ when both the modeled noise and the true map noise are $10\,\ukarcmin$, and also the case when the detector noise is $30\%$ higher (in map space) than modeled in the fixed $\mat{N}$ term. 

A simple extension to the covariance model can also fit for the detector noise amplitude in the maps with the covariance model, as \citep{2009ApJ...702L..87G} 
\begin{equation}
\mat{C}(\{r, s, \sigma \}) = r \tcov + s \scov + (\sigma / \sigma_0)^2 \mat{N}.
\label{eqn:simplecovnoise}
\end{equation}
Figure\,\ref{fig:detnoise_fit} shows that the distribution of $r$ is correctly centered around the input $r=0.05$ when the noise amplitude of $13\,\ukarcmin$ is jointly modeled with the cosmological signal as in Eq.\,(\ref{eqn:simplecovnoise}). It is slightly broader due to the higher level of detector noise and expense of fitting $\sigma$.

It is possible to fit for cosmological amplitudes $r$ and $s$ in parallel with noise $\sigma$ because these have different covariance structures. In multipole space, $s$ modulates the $E$-modes, $r$ impacts the $E$- and \bmodes\ and detector noise contributes to both but with a different $\ell$ dependence. When a contaminant has more overlap with the \bmode\ covariance structure, it becomes harder to separate. In the worst case, the excess covariance structure is identical to the signal and so is indistinguishable. 

\begin{figure}[htb]
\includegraphics[scale=0.55]{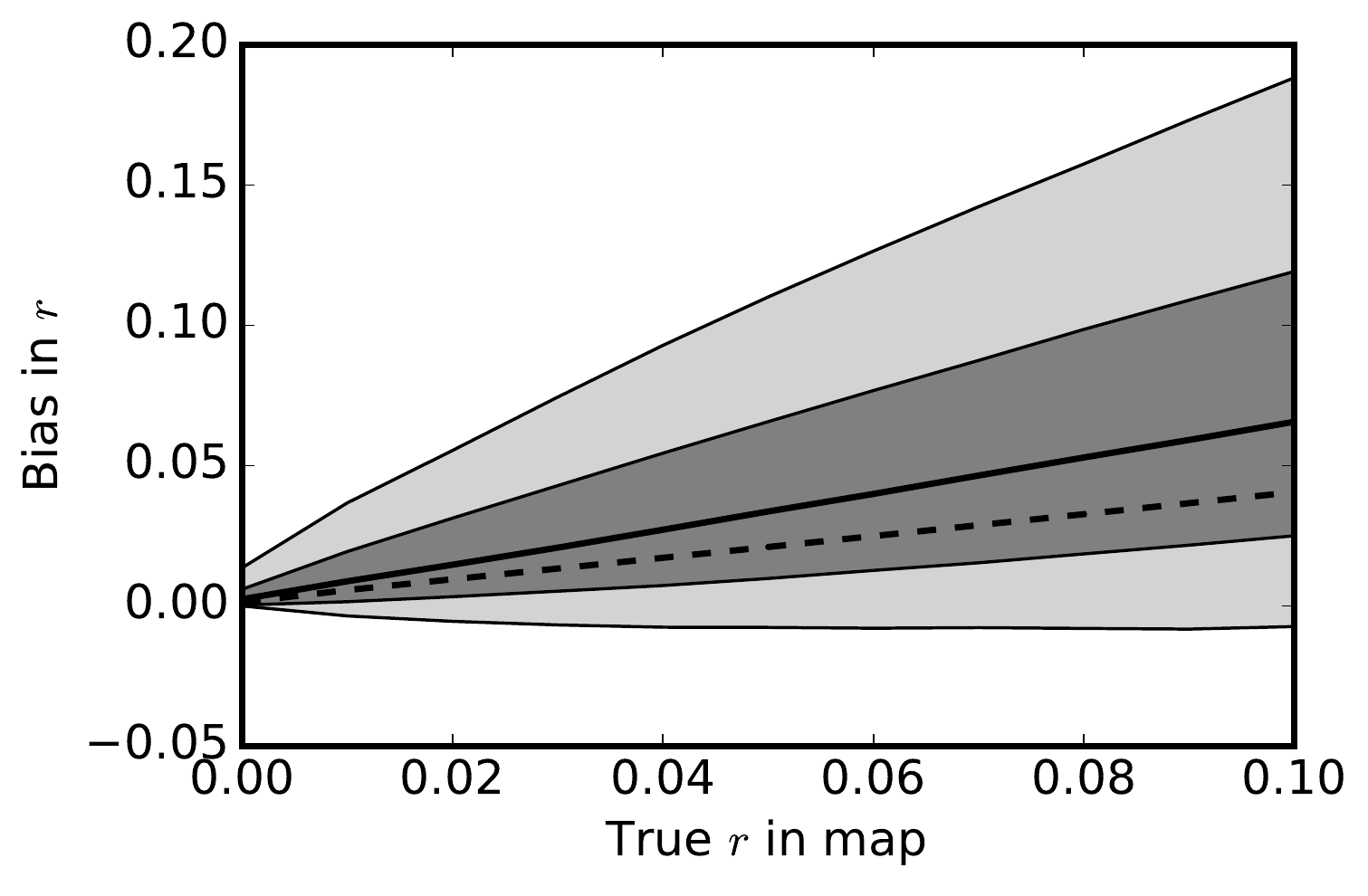}
\caption{The bias in $\hat r_{\rm ML}$ from detector noise that is $30\%$ higher than modeled in the likelihood, as a function of the true \bmode\ $r$ signal in the map. Light and dark gray regions are percentile regions equivalent to $2\sigma$ and $1\sigma$ respectively, about the median (solid line). The distribution is estimated as the Monte Carlo of the maximum likelihood value of $r$ over map realizations of both signal and detector noise. The dashed line shows bias in $r$ inferred from the quadratic estimate of $r$, as the first step of the Newton-Raphson approach to the maximum likelihood (Sec.\,\ref{ssec:likequad}). The presence of true \bmode\ signal modulates the weighting of variance that contributes to the bias.
\label{fig:rbias}}
\end{figure}

Figure\,\ref{fig:rbias} shows the $1\sigma$ and $2\sigma$ equivalent regions for the distribution of bias in $\hat r_{\rm ML}$ that is produced by detector noise $30\%$ higher than modeled. The bias has clear dependence on the amplitude of true \bmodes\ present in the map. The parameter dependence of the bias goes against intuition from the power spectrum, where uncorrelated contributions in the map are additive in $C_\ell$. The behavior in Fig.\,\ref{fig:rbias} does not occur for all types of covariance. For example, adding $r=0.1$ \bmodes\ to a map with $r=0$ \bmode\ amplitude will produce an estimate of $\langle \hat r \rangle = 0.1$. Adding a systematic with equivalent $r=0.1$ \bmodes\ to a map with true signal $r=0.1$ \bmode\ amplitude produces an estimate of $\langle \hat r \rangle = 0.2$. More generally, for a systematic level of $r_{\rm sys}$, the measured $\langle \hat r \rangle = r_{\rm true} + r_{\rm sys}$, independently of the level of $r_{\rm true}$. 

\subsection{Parameter dependence of the bias}
\label{ssec:bbias}

There is no intuitive closed-form expression for the maximum likelihood, but the quadratic approach to the maximum likelihood can give approximate expressions that explain the basic behavior. The quadratic estimate for $\hat r$ in Eq.\,(\ref{eqn:simplecov}) is analogous to the covariance amplitude determination in Eq.\,(\ref{eqn:quadest}), and gives
\begin{eqnarray}
\hat r &=& (\vect{x}^T \mat{C}^{-1} \tcov \mat{C}^{-1} \vect{x} - b) / \Tr(\mat{C}^{-1} \tcov \mat{C}^{-1} \tcov) \\
b &=& \Tr(\mat{C}^{-1} \tcov \mat{C}^{-1} \mat{N}),
\end{eqnarray}
where matrices are as defined in Eq.\,(\ref{eqn:simplecov}).
Taking the expectation value gives $\Tr(\mat{C}^{-1} \tcov \mat{C}^{-1} \langle \vect{x} \vect{x}^T \rangle)$ in the numerator. Identify the total covariance 
\begin{equation}
\mat{C}_{\rm tot} \equiv \langle \vect{x} \vect{x}^T \rangle = r \tcov + s \scov + \mat{N} + \mat{\Sigma},
\end{equation}
where $\mat{\Sigma}$ is some contamination covariance present in the data but not the model. Thermal noise is represented by $\mat{N}$ in the model, and is removed through subtraction of $b$. The remaining incurred bias in $\langle \hat r \rangle$ due to the un-modeled term $\mat{\Sigma}$ is
\begin{equation}
\langle \hat r_{\rm bias} \rangle = \Tr(\mat{C}^{-1} \tcov \mat{C}^{-1} \mat{\Sigma}) / \Tr(\mat{C}^{-1} \tcov \mat{C}^{-1} \tcov).
\label{eqn:rbiasquad}
\end{equation}
In the numerator, $\mat{C}^{-1} \mat{\Sigma}$ weighs the contamination $\mat{\Sigma}$, and $\mat{C}$ depends on the cosmological parameters $r$ and $s$ through covariance terms $r \tcov + s \scov$. Hence the weighting of the contamination is parameter dependent. Other parameters such as $\langle \hat s_{\rm bias} \rangle$ have a similar expression of some weighted overlap of the signal covariance structure with the contamination. The contamination biases all parameters that have structural overlap, and in a way that depends on the value of the parameters. Figure\,\ref{fig:rbias} shows the quadratic approximation to the bias as a dashed line. It is lower than the median of maximum likelihood realizations but captures much of the effect. Recall that the quadratic expression is only the first step of the Newton-Raphson approach to the maximum likelihood.

\begin{figure}[htb]
\includegraphics[scale=0.55]{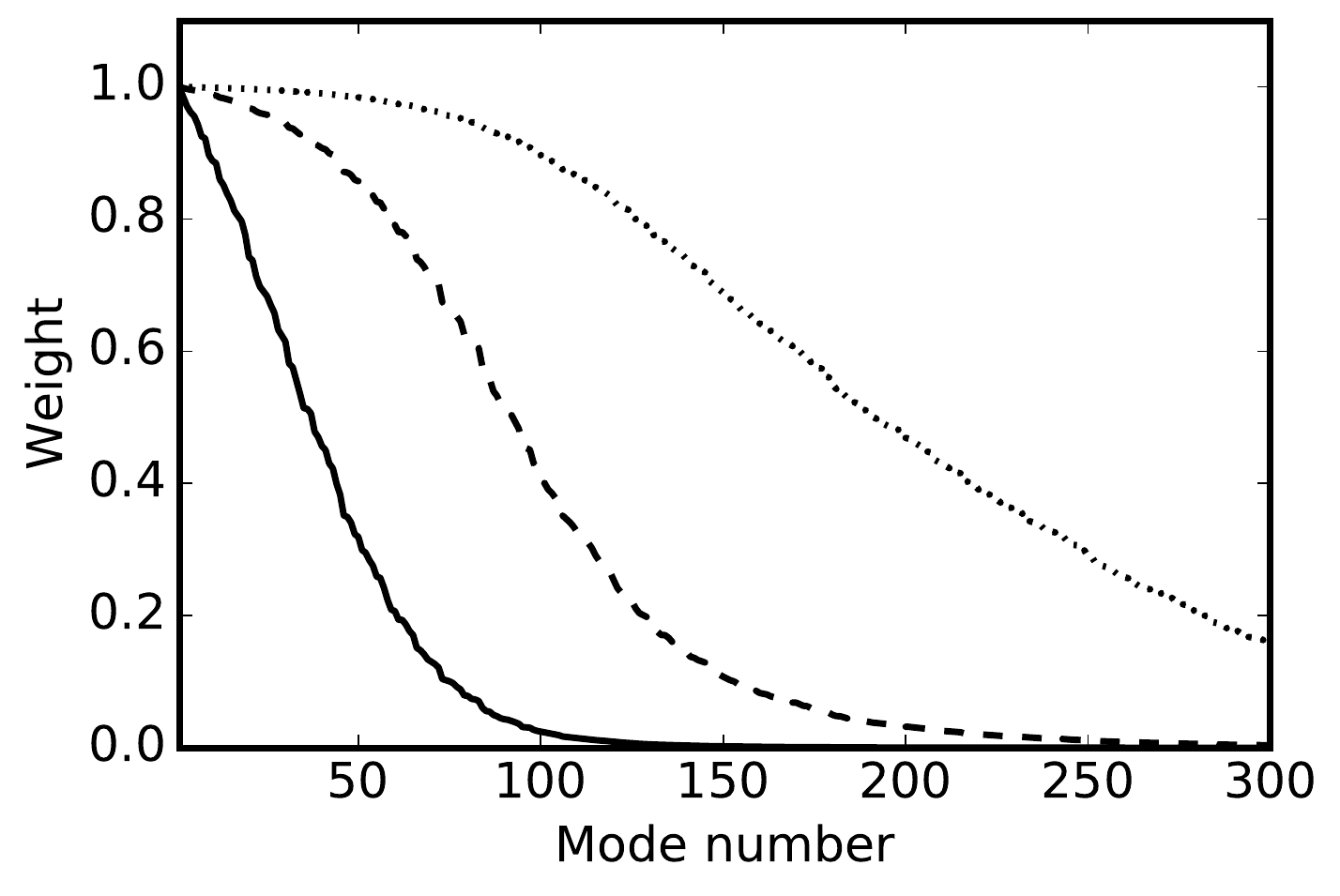}
\caption{$S/(S+N)$ weight per eigenmode for three values of tensor amplitude: $r=0.01$ (solid curve), $r=0.1$ (dashed curve), and $r=1$ (dotted curve). The parameter dependence of the weight produces parameter dependence in the bias from contamination.
\label{fig:bias_basis}}
\end{figure}

To get additional analytic intuition, Appendix~\ref{app:quadweight} derives $\langle \hat r_{\rm bias} \rangle$ in terms of the signal and contaminant covariance eigenvalues in the case where the contaminant is a multiplier times the identity matrix, $\mat{\Sigma} = \sigma_b^2 \mat{1}$. Then the bias becomes a simple ratio
\begin{equation}
\langle \hat r_{\rm bias} \rangle = \sigma_b^2 \frac{\vect{w}(r)^T \vect{1} }{\vect{w}(r)^T \vect{\lambda}},
\label{eqn:quadbiasweight}
\end{equation}
where $\vect{w}(r)$ is the weight per signal eigenmode (given in Appendix~\ref{app:quadweight}), and $\vect{\lambda}$ is the vector of signal eigenvalues. Figure\,\ref{fig:bias_basis} shows the terms of $\vect{w}(r)^T \vect{\lambda}$ for each mode, as a function of $r$, and is $S/(S+N)$ per mode. For low $r$, there are fewer signal-dominated modes. The $r$-dependence of the weight gives $r$-dependence to the bias. If the contaminant has the same form as the \bmode, then the numerator becomes $\vect{w}(r)^T \vect{\lambda}$, and the bias is independent of any true $r$ in the map. 

The quadratic estimator in Eq.\,(\ref{eqn:quadest}) has additional flexibility that is not available in the likelihood because it separates the covariance structure of the signal from the covariance structure of the weights. That is, in the quadratic form 
\begin{equation}
\hat \theta \propto \vect{x}^T \mat{\tilde C}^{-1} \mat{C}_{,\theta} \mat{\tilde C}^{-1} \vect{x},
\end{equation}
$\mat{\tilde  C}^{-1}$ and $\mat{C}_{,\theta}$ do not need to originate from the same parent $\mat{C}$. In the likelihood, this choice is fixed for both. Different choices of weights affect the optimality of the estimator and the structure of parameter correlations. In the pseudo-$C_\ell$ approach to spectral estimation, the quadratic combination of data can be weighted by some $\mat{N}^{-1}$ (hit map) or other apodization $\mat{W}$. In either case, the weighting does not depend on the parameters. 

\section{Likelihoods across data splits}
\label{sec:jointlikelihood}

Many sources of contamination vary or become uncorrelated across subsets of data from different epochs, frequencies, surveys, or data treatments. The cross power estimator \citep{2005MNRAS.358..833T} extends the quadratic estimator in Eq.\,(\ref{eqn:quadest}) by forming the quadratic product across subsets $A$ and $B$ of the data as
\begin{equation}
\hat C_\ell \propto \vect{x}_A^T \mat{N}_A^{-1} \mat{S}_{,\ell} \mat{N}_B^{-1} \vect{x}_B.
\label{eqn:quadcross}
\end{equation}
The expectation value $\langle \hat C_\ell \rangle \propto \Tr(\mat{N}_A^{-1} \mat{S}_{,\ell} \mat{N}_B^{-1} \langle \vect{x}_B \vect{x}_A^T \rangle)$ contains $\vect{x}_B \vect{x}_A^T$, which averages to zero for any variance terms that are not common to both $A$ and $B$. (Variance not common between $A$ and $B$ increases the variance of the estimator.) This approach has recently been extended to analysis of variance when few modes are available in a survey volume \citep{2015MNRAS.453.3174M}. 

The difference of maps across a data split, $\vect{x}_A - \vect{x}_B$, will remove any astronomical signal common to both maps. The power spectrum of the difference map tests for any excess variance. For example, the analysis of time-domain data must account or compensate for the detector response time constants. Otherwise, time constants can produce a residual variance in the difference between maps of left- and right-going scans. Large suites of such null tests support the ultimate parameter determination by ruling out sources of contamination. This approach has been applied extensively to spectral analysis and, to a much more limited degree, pixel-space likelihoods \citep{2015arXiv150702704P}. 

\subsection{The joint likelihood}
\label{sec:jointl}

The pixel-space likelihood analogy to the cross power is the joint likelihood of the maps $A$ and $B$ in the data split. Model the joint likelihood across the data split between $A$ and $B$, with $\vect{x}^T = (\vect{x}_A^T~~\vect{x}_B^T)$ as 
\begin{equation}
\mat{C} = \left ( \begin{array}{cc} \mat{S}(\vect{\Theta}) + \mat{N}_A & \mat{S}(\vect{\Theta}) \\ \mat{S}(\vect{\Theta}) & \mat{S}(\vect{\Theta}) + \mat{N}_B \end{array} \right),
\label{eq:jointcovsimple}
\end{equation}
where $\mat{S}(\vect{\Theta})$ is in common to both $A$ and $B$. Recall that $\vect{x}_A$ is still a stack of the Stokes $Q$ and $U$ maps, so the combined data vector $(\vect{x}_A^T~~\vect{x}_B^T)$ is the stack of four maps and the covariance is also naturally $4\times 4$ blocks for correlations of Stokes $Q$ and $U$ across $A$ and $B$. The noise covariance can also be extended to accept parameters such as the $\sigma$ amplitude in Sec.\,\ref{sec:detnoise}.  

A likelihood model for that data that uses Eq.\,(\ref{eq:jointcovsimple}) extracts parameter information from $A \times B$ (cross) but also $A \times A$ and $B \times B$ (auto). This can be seen in the form of the quadratic estimator $\vect{x}^T \mat{C}^{-1} \mat{C}_{,\theta} \mat{C}^{-1} \vect{x}$, where both $\mat{C}^{-1}$ and $\mat{C}_{,\theta}$ have off- and on-diagonal block terms, so the inner product with $\vect{x}^T = (\vect{x}_A^T~~\vect{x}_B^T)$ mixes both $A\times B$ and $A \times A$ or $B \times B$ information. A likelihood that uses the covariance model of Eq.\,(\ref{eq:jointcovsimple}) will have sensitivity to information in $A \times A$, and so does not have the same immunity to uncorrelated noise as the cross power in Eq.\,(\ref{eqn:quadcross}).

To reach a closer analog to the cross power, take a covariance model which has a duplicate set of nuisance parameters on the diagonal as
\begin{equation}
\mat{C} = \left ( \begin{array}{cc} \mat{S}(\vect{\Theta}) + \mat{S} (\vect{\Theta}_{b}) + \mat{N}_A & \mat{S}(\vect{\Theta}) \\ \mat{S}(\vect{\Theta}) & \mat{S}(\vect{\Theta}) + \mat{S}(\vect{\Theta}_{b}) + \mat{N}_B \end{array} \right).
\label{eq:jointcov}
\end{equation}
Adding $\mat{S} (\vect{\Theta}_{b})$ to the diagonal and marginalizing over $\vect{\Theta}_{b}$ effectively sweeps the rug out from under parameter constraints on $\vect{\Theta}$ coming from $A \times A$ and $B\times B$.
Appendix~\ref{app:jointcross} relates Eq.\,(\ref{eq:jointcov}) to the cross quadratic product $\vect{x}_A^T \mat{N}_A^{-1} \mat{C}_{,\theta} \mat{N}_B^{-1} \vect{x}_B$ (with no noise bias to remove) as the first step of a Newton-Raphson iteration starting from a noise-only covariance.

The approach in Eq.\,(\ref{eq:jointcov}) resembles mode avoidance or cleaning strategies. These are commonly implemented by fitting and subtracting mode functions. Several authors \citep{1992ApJ...398..169R, 1998ApJ...499..555T, 1998ApJ...503..492S, 2010MNRAS.408..865T} consider this class of avoidance and argue that the following three are equivalent: (1) least-squared fitting and subtracting modes in the mean model, (2) marginalizing over the amplitude of nuisance modes in the mean model, and (3) taking a multiplier of the covariance structure of the contaminated modes to infinity. In the present case, rather than marginalizing over the bias nuisance variables $\vect{\Theta}_{b}$, another approach that suggests itself would be to set the amplitude of these on-diagonal variance terms to infinity through $\vect{\Theta}_{b}$. This limit throws away dependence on the nuisance parameters, and so halves the number of free parameters to estimate through the likelihood. Taking on-diagonal signal variance to infinity results in a quadratic estimator that uses only information across the data split, as intended. However, it also results in a $\mat{C}^{-1}$ weight applied to each map with infinite variance in the signal modes; e.g., it also eliminates the signal. Rather than marginalize over nuisance parameters in the mean, the likelihood using the covariance model in Eq.\,(\ref{eq:jointcov}) marginalizes over nuisance parameters in the covariance. \citet{2010MNRAS.408..865T} give analytic expressions for marginalization over covariance parameters. These can be used to simplify computation in a high-dimensionality parameter space, but are not needed for the simple, few-parameter models considered here.

\subsection{The sum-difference likelihood}
\label{ssec:sumdifflike}

Rotating to a basis of sum and difference maps, $(\vect{x}_s, \vect{x}_d) = (\vect{x}_A^T + \vect{x}_B^T, \vect{x}_A^T - \vect{x}_B^T)$ simplifies the joint covariance in Eq.\,(\ref{eq:jointcov}). Take the detector noise amplitude to be identical between splits $A$ and $B$ for simplicity (this can be arranged by splitting complementary sets of the data with common integration depth). The resulting likelihood separates as the product of sum and difference likelihoods
\begin{eqnarray}
\label{eqn:sumdiff}
P(\vect{x} | \vect{\Theta},  \vect{\Theta}_b) &=& P_s(\vect{x}_s | \vect{\Theta},  \vect{\Theta}_b) P_d(\vect{x}_d | \vect{\Theta}_b) \\
P_s(\vect{x}_s | \vect{\Theta},  \vect{\Theta}_b) &\sim& N(\vect{0}, 4 \mat{S}(\vect{\Theta}) + 2 \mat{S} (\vect{\Theta}_{b}) + 2 \mat{N}) \nonumber \\
P_d(\vect{x}_d | \vect{\Theta}_b) &\sim& N(\vect{0}, 2 \mat{S}(\vect{\Theta}_{b}) + 2 \mat{N}). \nonumber 
\end{eqnarray}
The factors of $2$ and $4$ are a by-product of taking the $\vect{x}_A^T + \vect{x}_B^T$ combination rather than $1/2 \cdot (\vect{x}_A^T + \vect{x}_B^T)$. The signal in common to both maps appears as $4 \mat{S}(\vect{\Theta})$ because it adds coherently between the two maps, as  $(2 \sigma)^2$, and the bias of uncorrelated contamination appears in both the sum and the difference as $2 \mat{S} (\vect{\Theta}_{b})$ because it is the addition of two uncorrelated variances $\mat{S} (\vect{\Theta}_{b}) + \mat{S} (\vect{\Theta}_{b})$ in each map. 

An interpretation of the joint sum-diff likelihood of Eq.\,(\ref{eqn:sumdiff}) is that $P_s$ constrains the parameters plus bias, while the difference $P_d$ constrains only bias from modulated contamination. 

\subsection{Sum-difference likelihood: \bmode\ contamination}
\label{ssec:sdknown}

\begin{figure}[htb]
\includegraphics[scale=0.7]{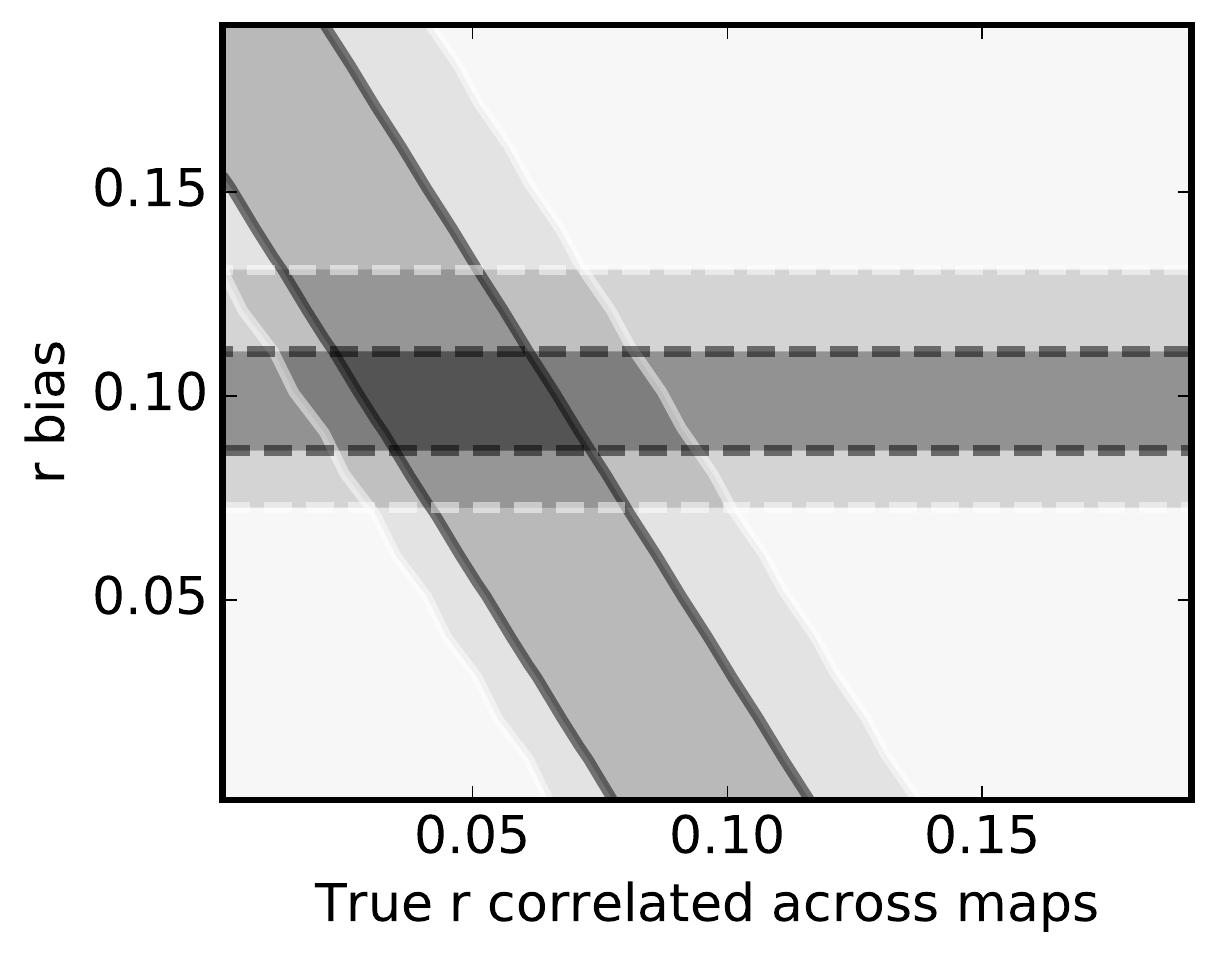}
\caption{The likelihood of $r_{\rm bias}$ and $r_{\rm true}$ from the sum and difference of maps across a split in the data. Here $r_{\rm true}=0.05$ is in common between the maps, and each map has an independent realization of contaminating \bmodes\ at the level $r_{\rm bias} = 0.1$. This represents a scenario where there is time-varying contamination with structure indistinguishable from the \bmodes. {\it Solid lines:} The regions of $68\%$ and $95\%$ probability enclosed ($1\sigma$, $2\sigma$ equivalent) in the likelihood of the sum map. It traces a degeneracy between interpretation of the data as all true \bmode\ signal or all bias. {\it Dashed lines:} The analogous regions for the likelihood of the difference map. This likelihood isolates residual variance at the level $r_{\rm bias} = 0.1$. The joint likelihood of sum and difference maps determines both the bias and the target $r$ amplitude with self-consistent errors. Both posterior distributions are from a single realization of data, so the maximum likelihood is not centered on the input parameters.
\label{fig:sd_joint}}
\end{figure}

Following Sec.\,\ref{sec:detnoise}, take a simple model where $\mat{S}(\vect{\Theta}) = r_{\rm true} \tcov + s_{\rm true} \scov$. Again $\tcov$ and $\scov$ are the covariance structure of the tensor and scalar modes in the Stokes $Q$ and $U$ maps. The new parameters in the bias space are $\mat{S}(\vect{\Theta}_b) = r_{\rm bias} \tcov + s_{\rm bias} \scov$. Figure\,\ref{fig:sd_joint} shows the sum and difference likelihoods for $r$ for a simulation with $r_{\rm true} = 0.05$, detector noise of $10\,\ukarcmin$ in each map, and contamination at the level of $r_{\rm bias} = 0.1$ in uncorrelated realizations added to both maps. The sum map can only constrain $2 r_{\rm true} + r_{\rm bias}$ so it is a degenerate band from upper left to lower right. The difference map can only constrain $r_{\rm bias}$ with no dependence on $r_{\rm true}$, so it appears as a horizontal band. The product of the two likelihoods recovers both the input $r_{\rm true} = 0.05$ and the contamination level. The factor of two in $2 r_{\rm true} + r_{\rm bias}$ is effectively a normalization for the bias amplitude under the assumption that the source of bias is uncorrelated between the split maps.

Marginalizing over $r_{\rm bias}$ in the joint likelihood gives an estimate of $r_{\rm true}$ with errors self-consistently inflated to reflect the fact that some of the constraining power of the map is used to estimate bias. An analogous power-spectral null test uses the difference of maps to rule out bias parameters. If the power spectrum of the difference map is consistent with zero, a particular source of contamination can be ruled out. In a typical use of the null test, once a source of variance is ruled out, it is taken to be identically zero. In contrast, in the joint likelihood, Eqs.\,(\ref{eqn:sumdiff}) and\,(\ref{eq:jointcov}), uncertainty in the bias parameter is folded into the final estimate $\hat r_{\rm true}$.

When the likelihood model includes all covariance terms, the tensor to scalar ratio can be recovered without bias and with correct confidence intervals. When the contaminant covariance differs in structure from terms in the covariance model, the cross likelihood fails and can give biased results due to the weighting effect in Sec.\,\ref{ssec:bbias}. 

\subsection{Reweighting for difference map null tests}
\label{ssec:rewighted}

Section~\ref{sec:detnoise} provides a scenario where the data have some covariance that is not explained by the structure in the likelihood model. In the case of a single map, Sec.\,\ref{ssec:bbias} showed how a parameter $\theta$ is biased when its structure $\mat{C}_{,\theta}$ overlaps with contamination $\mat{\Sigma}$ through $\Tr(\mat{C}^{-1} \mat{C}_{,\theta} \mat{C}^{-1} \mat{\Sigma})$. Further, the bias depends on parameters through the $\mat{C}(\vect{\Theta})^{-1}$ weight. In the sum-difference formalism, the difference map has no signal by construction, so the contamination is weighted differently than in the sum map likelihood. 

Mis-specification of the covariance model results in significantly different parameter biases in the likelihoods of the sum and difference maps. In the difference map, there is no cosmological \bmode\ signal by construction, so the pixel-space covariance is equivalent to the case of $r=0$. Figure\,\ref{fig:rbias} shows the bias produced by a $30\%$ excess of thermal noise as a function of $r$ assumed in the covariance. At $r=0$, the likelihood of the difference map reports negligible bias produced by the excess thermal noise. In contrast, if the data have a true $r=0.1$, the $30\%$ excess thermal noise will produce a bias of $\Delta r = 0.06$ on average, with fluctuations at the level of $\sigma_r = 0.05$. The parameters inferred from the likelihood of the difference map no longer provide useful information about the bias of parameters in the sum map, and may lead to false confidence in the analysis.

For the likelihood of the difference to constitute a null test, it must weigh the contamination consistently with the sum. An approach to consistent weighting is to add a signal realization $\vect{x}_{\rm sig}$ to the difference map $\vect{x}_d$ and find the Monte Carlo average of the parameters over signal realizations (each realization will have some signal variance). Any deviation from the input parameters could be attributed to contamination in the difference map. In this case, the data matrix $\langle \mat{D} \rangle = \langle (\vect{x}_{\rm sig} + \vect{x}_d) (\vect{x}_{\rm sig} + \vect{x}_d)^T \rangle = \vect{x}_d \vect{x}_d^T + 4 \mat{S} (\vect{\Theta})$, where $\langle \rangle$ is the expectation over signal realizations. (In the sum map, two coherent copies of the signal are added, giving a factor of $4$ in variance.) Rather than Monte Carlo, we take the expectation value of the likelihood over added signal, giving the model
\begin{eqnarray}
2 \mathcal L_d &=& \Tr[\ln \mat{\tilde C}_d + \mat{\tilde C}_d^{-1} (\vect{x}_d \vect{x}_d^T + 4 \mat{S}(\vect{\Theta})) ] \label{eqn:diffreweightlike} \\
\mat{\tilde C}_d &=& 2 \mat{S}(\vect{\Theta}_{b}) + 2 \mat{N} + 4 \mat{S}(\vect{\Theta}). \nonumber
\end{eqnarray}
The factor of 2 in the bias covariance $2 \mat{S}(\vect{\Theta}_{b})$ represents the assumption that the excess variance producing the systematic is not correlated across the difference of maps.

The role of $4 \mat{S}(\vect{\Theta})$ in the covariance model is clear as a reweighting, but the $4 \mat{S}(\vect{\Theta})$ term in the data matrix also plays an important role in the likelihood. In the Newton-Raphson approach to maximum likelihood, each iteration is based on the difference between the data matrix (outer product of the data) and the covariance model,
\begin{eqnarray}
\mat{\Delta} &=& \mat{D} - \mat{C} (\vect{\Theta}) \\
&=& [\vect{x}_d \vect{x}_d^T + 4 \mat{S}(\vect{\Theta})] - [2 \mat{S}(\vect{\Theta}_{b}) + 2 \mat{N} + 4 \mat{S}(\vect{\Theta})] \\
&=& \vect{x}_d \vect{x}_d^T - 2 \mat{S}(\vect{\Theta}_{b}) - 2 \mat{N}.
\end{eqnarray} 
The maximum likelihood therefore fits the residual variance in the difference map to the signal bias model, accounting for thermal noise. Recall that each Newton-Raphson step in Eq.\,(\ref{eq:nriter}) is weighted by a $\mat{C}^{-1}$, which also contains $4 \mat{S}(\vect{\Theta})$ and weights $\mat{\Delta}$ consistently with the sum map.

The reweighted likelihood of the difference map should be interpreted as $P_d(\vect{\Theta}_b | \vect{\Theta})$, the distribution of bias parameters evaluated in a map where signal variance is fixed at $\vect{\Theta}$. It should not be interpreted as the joint likelihood $P_d(\vect{\Theta}_b, \vect{\Theta})$.

The sum and difference likelihoods can be sampled and combined through the following process:
\begin{enumerate}
\item Use the map sum $\vect{x}_A + \vect{x}_B$ to constrain $2 r_{\rm true} + r_{\rm bias}$, marginalized over all other parameters. This gives a diagonal band of degeneracy in the $r_{\rm true}$-$r_{\rm bias}$ plane.
\item Use the difference map $\vect{x}_A - \vect{x}_B$ in the re-weighted likelihood Eq.\,(\ref{eqn:diffreweightlike}) ($r_{\rm true}$ fixed) to estimate the distribution of $r_{\rm bias}$, marginalized over all other parameters. This gives a slice of the probability of $r_{\rm bias}$ for a given true level of signal $r_{\rm true}$, $P(r_{\rm bias} | r_{\rm true})$.
\item Repeat the difference analysis for any other null test combinations, giving contours in the $r_{\rm true}$-$r_{\rm bias}$ plane. 
\end{enumerate}

\begin{figure}[htb]
\includegraphics[scale=0.70]{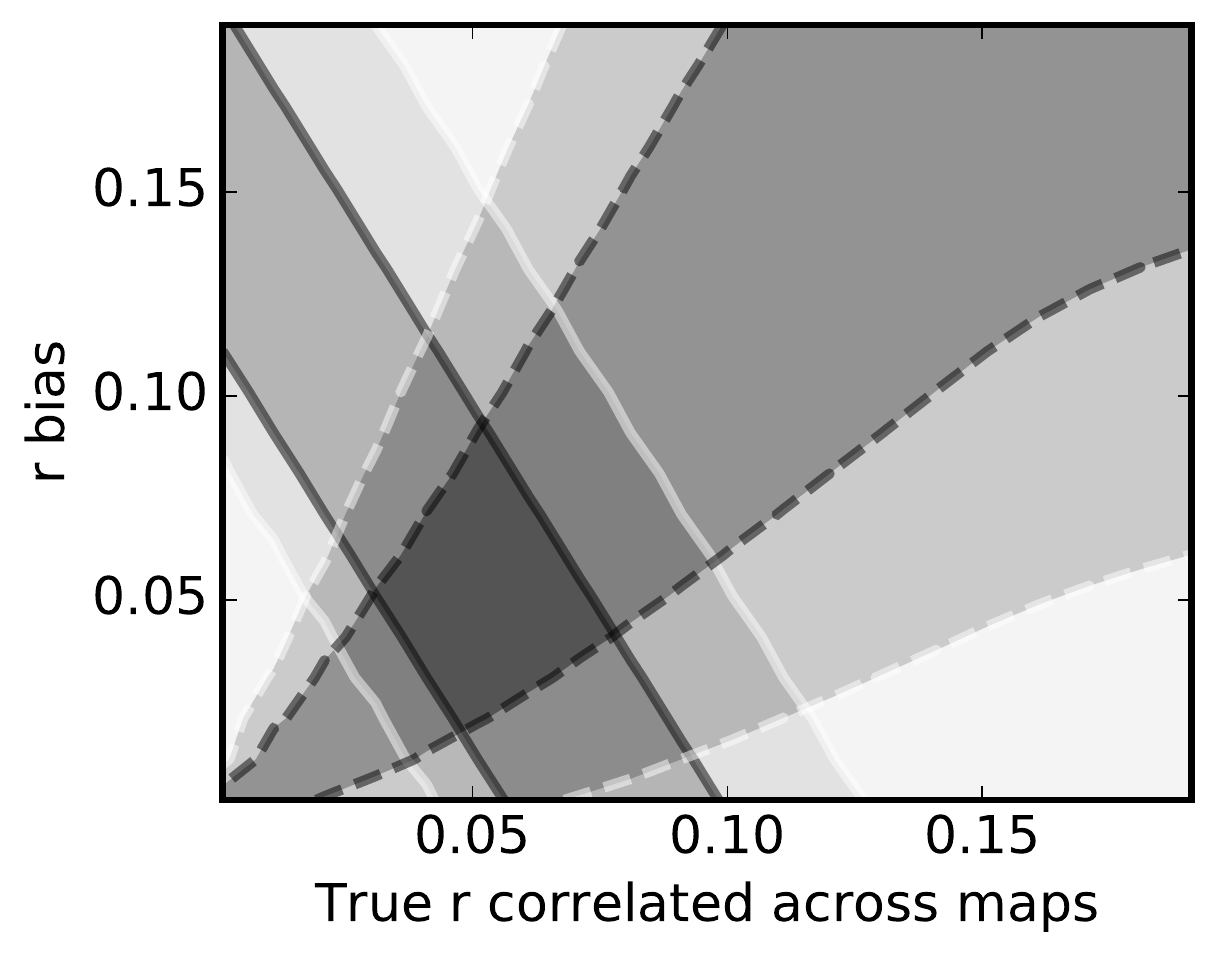}
\caption{Same layout as Fig.\,\ref{fig:sd_joint}, except that instead of adding variance in the form of \bmodes\ (which are in the likelihood covariance model), this simulation has detector noise $30\%$ higher than modeled and is not explained by any free term of the covariance. Mis-specification produces a parameter-dependent bias. Here we force the likelihood of the difference map to weight consistently with the covariance in the sum map likelihood. Without reweighting, the likelihood of the difference map gives $r_{\rm bias} < 0.004$ at $95\%$ confidence.
\label{fig:sd_reweight}}
\end{figure}

Figure\,\ref{fig:sd_reweight} applies this process to maps with $r_{\rm true}=0.05$ and detector noise that is $30\%$ higher than is modeled in the likelihood. Without reweighting, the likelihood of the difference map reports $r_{\rm bias} < 0.004$ at $95\%$, independently of $r_{\rm true}$. Reweighting the likelihood results in $P(r_{\rm bias} | r_{\rm true})$ that depends on $r_{\rm true}$, analogously to Fig.\,\ref{fig:rbias}. 

WMAP \citep{2009ApJS..180..225H} and Planck \citep{2015arXiv150702704P} find the posterior distribution of $\tau$ from difference maps, but do not describe added signal covariance in the model. This covariance is required to consistently weight contamination in the null analysis.

To constitute a useful null test, the likelihood of the difference map must also give an informative confidence interval. Appendix~\ref{app:likewidth} describes the curvature of the posterior parameter distribution. When there is no excess variance, the curvature is the usual Fisher matrix $\Tr [\mat{C}^{-1} \mat{C}_{,\theta_i} \mat{C}^{-1} \mat{C}_{,\theta_j}]$. The width coincides with the distribution of the cosmological signal and noise at fixed contamination. The width is analogous to a standard null test, where the errors in the difference map power spectrum do not account for any contamination. 

The width of the likelihood is erroneous when the difference map has an excess variance that is not described by free parameters in the likelihood model. Contamination must be treated either in a reprocessing of the data or be modeled self-consistently in the likelihood. Examples here could include a fit for noise amplitude in the map or deprojection \citep{2015ApJ...811..126B}, where instrumental systematics of known structure produce correlations between temperature and polarization maps. Extensions to the likelihood must balance adequacy (of describing noncosmological variance) and simplicity. The likelihood ratio and related tests can be used to assess the candidate extensions.

\section{Summary}
\label{sec:summary}

Instrumental systematics, residual foregrounds, and other excess variance produce bias in cosmological parameters. Experiments to detect inflationary gravitational waves must use a battery of tests to rule out biases that could lead to a false detection. Determination of cosmological parameters directly from pixel-space likelihood has shown promise as a method to self-consistently handle foregrounds and survey depth variations or masks, especially on the largest scales in the survey. This approach bypasses calculation of the band powers, which have had a vigorous history of tests for systematics. We have developed some pixel-space likelihood analogies to the cross power, noise modeling, and the difference map null test.

If excess variance modulates with time or instrumental setup, a difference map can be interpreted using a likelihood for a particular source of parameter bias. The two dimensional posterior distribution of a parameter and its bias is a convenient diagnostic. We show examples of this parameter bias plane for the tensor to scalar ratio $r$ when the excess variance is parameterized in the likelihood, and where it is not (mis-specification).

The parametric nature of the likelihood requires additional care. Bias in the pixel-space likelihood is signal-dependent because the map weights contain signal covariance. Signal dependence of the weight produces parameter distributions that are not consistent between the sum and difference maps. We propose a procedure for consistently weighting contamination. The method accomplishes the role of a null test: under the same assumptions as the sum map analysis, is there evidence for parameter bias caused by modulated contamination in the difference?

We recommend an iterative approach. In a first pass, the likelihood models all cosmological parameters and imperfectly known instrumental terms (e.g. detector noise). If a weighted difference null test fails, that information should be used either to construct a model of excess variance or to reprocess the data in a way that eliminates the systematic effect. If a left-right scan difference fails, compensation of time constants should be reassessed until that test passes. Temperature-to-polarization leakage results in a covariance matrix between the temperature and the polarization. In parallel with cosmological parameters, the pixel-space likelihood should include any contamination which has a well-defined model.

The likelihood of the difference map provides a parametric test for mis-specification of the covariance model by isolating components that vary across data subsets. A more general problem is assessing whether there is variance in the data, time-varying or not, that is not explained well by the model and could produce spurious \bmodes. A parametric model can be tested against less parametric models that are sensitive to a wider range of variance structure. In the case of CMB polarization, the power spectrum is already an excellent example of this approach and has been used by WMAP \citep{2009ApJS..180..225H} and Planck \citep{2015arXiv150702704P} to corroborate likelihood results on large scales. The power spectrum exposes statistically isotropic variance with $\ell$-dependence different from the signal. At the next level, tests for isotropy \citep{2014A&A...571A..23P, 2014PhRvL.113s1303K} are sensitive to residual galactic foregrounds. 

\acknowledgements 

We acknowledge useful comments from Jo Dunkley and an anonymous reviewer.

\appendix
\onecolumngrid

\section{Relation of the joint likelihood and cross power approaches}
\label{app:jointcross}

For a model with only \bmodes, $\vect{\Theta} = \{ r, r_b\}$, the joint covariance across subseason maps $A$ and $B$ is
\begin{equation}
\mat{C} = \left( \begin{array}{cc} (r + r_b) \tcov + \mat{N}_A & r \tcov \\ r \tcov & (r + r_b) \tcov + \mat{N}_B \end{array} \right)
\end{equation}

As a first iteration of the Newton-Raphson (NR) approach in Eq.\,(\ref{eq:nriter}), take the case where $r = r_b = 0$. In this case,
\begin{equation}
\mat{F} = \left ( \begin{array}{cc} \alpha & \beta \\ \beta & \beta \end{array} \right)~~~~\alpha = \frac{1}{2} \Tr[ (\mat{N}_A^{-1} + \mat{N}_B^{-1}) \tcov (\mat{N}_A^{-1} + \mat{N}_B^{-1}) \tcov]~~~~
\beta = \frac{1}{2} \Tr[ \mat{N}_A^{-1} \tcov \mat{N}_A^{-1} \tcov + \mat{N}_B^{-1} \tcov \mat{N}_B^{-1} \tcov]
\end{equation}
and the initial iteration for $r$ is 
\begin{eqnarray}
\delta r = \frac{\Tr \left [ (\vect{x} \vect{x}^T -\mat{C}) \mat{Q} \right ]}{2 \Tr(\mat{N}_A^{-1} \tcov \mat{N}_B^{-1} \tcov)}~~~~~~\mat{Q} \equiv \mat{C}^{-1} \left ( \begin{array}{cc} \mat{0} & \tcov \\ \tcov & \mat{0} \end{array} \right) \mat{C}^{-1} 
\end{eqnarray}
where we have identified the form of an optimal quadratic estimator $\mat{Q}$.

The noise bias term
\begin{equation}
\Tr[ \mat{C} \mat{Q}] = \Tr \left [ \left ( \begin{array}{cc} \mat{0} & \tcov \\ \tcov & \mat{0} \end{array} \right) \left ( \begin{array}{cc} \mat{N}_A^{-1} & \mat{0} \\ \mat{0} & \mat{N}_B^{-1} \end{array} \right) \right ] = 0.
\end{equation}
The signal estimator term
\begin{equation}
\Tr [ \vect{x} \vect{x}^T \mat{Q} ] = 2 \vect{x}_A^T \mat{N}_A^{-1} \tcov \mat{N}_B^{-1} \vect{x}_B.
\end{equation}
If we take the estimator $\hat r$ to be this first NR iteration, it has the form of a cross power
\begin{equation}
\hat r = \frac{\vect{x}_A^T \mat{N}_A^{-1} \tcov \mat{N}_B^{-1} \vect{x}_B}{\Tr(\mat{N}_A^{-1} \tcov \mat{N}_B^{-1} \tcov)}.
\end{equation}

\section{Parameter dependence of the bias}
\label{app:quadweight}

Section~\ref{ssec:bbias} argues that the bias in $r$ from contamination covariance $\mat{\Sigma}$ is $\langle r_{\rm bias} \rangle = \Tr(\mat{C}^{-1} \tcov \mat{C}^{-1} \mat{\Sigma}) / \Tr(\mat{C}^{-1} \tcov \mat{C}^{-1} \tcov)$. As a toy model to understand the behavior analytically, take the contaminant $\mat{\Sigma} = \sigma_b^2 \mat{1}$ and noise $\mat{N} = \sigma_n^2 \mat{1}$. This is equivalent to having uniform detector noise in excess of what is predicted. Expand the \bmode\ signal covariance $r \tcov = r \mat{U} \mat{\Lambda} \mat{U}^T$. The inverse covariance according to the Woodbury inverse is
\begin{eqnarray}
\mat{C}^{-1} &=& (r \tcov + \mat{N})^{-1} = \mat{N}^{-1} - \mat{N}^{-1} \mat{U} (r^{-1} \mat{\Lambda}^{-1} + \mat{U}^T \mat{N}^{-1} \mat{U})^{-1} \mat{U}^T \mat{N}^{-1} \\
&=& \sigma_N^{-2} \mat{U} [\mat{1} - \sigma_N^{-2} ( r^{-1} \mat{\Lambda}^{-1} + \sigma_N^{-2} \mat{1})^{-1} ] \mat{U}^T.
\end{eqnarray}
Taking $r\rightarrow 0$, $\mat{C}^{-1} =  \sigma_N^{-2} \mat{1}$. This is just the uniform detector noise weight. The quadratic estimator part of the bias trace is
\begin{eqnarray}
\mat{C}^{-1} \tcov \mat{C}^{-1} &=&  
 \sigma_N^{-4} \mat{U} [\mat{1} - \sigma_N^{-2} ( r^{-1} \mat{\Lambda}^{-1} + \sigma_N^{-2} \mat{1})^{-1} ] \mat{\Lambda} [\mat{1} - \sigma_N^{-2} ( r^{-1} \mat{\Lambda}^{-1} + \sigma_N^{-2} \mat{1})^{-1} ] \mat{U}^T =  \sigma_N^{-4} \mat{U} \mat{W} \mat{U}^T,
\end{eqnarray}
where we have identified the weighting term $\mat{W}$ as the combination of factors between $\mat{U}$ and $\mat{U}^T$. The weight is implemented as a vector multiplication as long as the contaminant is diagonalized by the same vectors $\mat{U}$ as the signal, which in this case is possible because we chose contaminant $\mat{\Sigma}= \sigma_b^2 \mat{1}$. Here,
\begin{equation}
\langle \hat r_{\rm bias} \rangle = \frac{\Tr(\mat{U} \mat{W} \mat{U}^T \sigma_b^2 \mat{1})}{\Tr(\mat{U} \mat{W} \mat{U}^T \mat{U} \mat{\Lambda} \mat{U}^T )} = \sigma_b^2 \frac{\vect{w}(r)^T \mat{1}}{\vect{w}(r)^T \vect{\lambda}},
\end{equation}
where we have used the cyclic property of the trace, used the orthonormality of the eigenvectors, and let $\vect{\lambda}$ and $\vect{w}$ be the diagonals of the signal eigenvalues $\mat{\Lambda}$ and weight $\mat{W}$. In contrast, if the contaminant has the same covariance structure as the signal, $\hat r_{\rm bias} = \sigma_b^2 \vect{w}(r)^T \vect{\lambda} /\vect{w}(r)^T \vect{\lambda} = \sigma_b^2$ with no dependence on $r$. Each term of the product $\vect{w}(r)^T \vect{\lambda}$ is the $\sim S/(S+N)$ of the signal mode. Modes in the map that contribute less signal to noise are downweighted. The denominator $\Tr(\mat{C}^{-1} \tcov \mat{C}^{-1} \tcov) \propto \vect{w}(r)^T \vect{\lambda}$ is the effective number of independent modes of information about $r$ in the map, and it is also equal to the Fisher matrix.

\section{Curvature of the likelihood under mis-specification}
\label{app:likewidth}

The general curvature of the log-likelihood is \citep{1997ApJ...480...22T}
\begin{eqnarray}
2 \mathcal{L}_{,ij} = \Tr[&-&\mat{C}^{-1} \mat{C}_{,i} \mat{C}^{-1} \mat{C}_{,j} + \mat{C}^{-1} \mat{C}_{,ij} + \mat{C}^{-1} (\mat{C}_{,i} \mat{C}^{-1} \mat{C}_{,j} + \mat{C}_{,j} \mat{C}^{-1} \mat{C}_{,i}) \mat{C}^{-1} \mat{D} 
 \\ &-& \mat{C}^{-1} (\mat{C}_{,i} \mat{C}^{-1} \mat{D}_{,j} + \mat{C}_{,j} \mat{C}^{-1} \mat{D}_{,i}) - \mat{C}^{-1} ( \mat{C}_{,ij} \mat{C}^{-1} \mat{D} - \mat{D}_{,ij})].
\end{eqnarray}
where $\mat{D} = (\vect{x} - \vect{\mu}) (\vect{x} - \vect{\mu})^T$. In the case that the covariance model $\mat{C}$ is the same as the covariance of the data $\langle \mat{D} \rangle = \mat{C}$, and there is no parameter dependence in the mean, one recovers the usual Fisher matrix (expectation of the log-curvature)
\begin{equation}
\mat{F}_{ij} = \frac{1}{2} \Tr (\mat{C}^{-1} \mat{C}_{,i} \mat{C}^{-1} \mat{C}_{,j}).
\end{equation}

With some unknown contamination $\vect{x}_c$, the expectation value of $\mat{D}$ is $\mat{C} + \vect{x}_c \vect{x}_c^T$. Generally the maximum-likelihood estimate of the covariance is biased by the presence of this contamination, or $\mat{C}_{\rm biased}$. Neglecting parameter dependence of the mean ($\mat{D}_{,i}$) and second derivatives of the covariance (applicable for the models here with $r \tcov$) gives
\begin{eqnarray}
2 \langle \mathcal{L}_{,ij} \rangle = \Tr[&-&\mat{C}^{-1} \mat{C}_{,i} \mat{C}^{-1} \mat{C}_{,j} + \mat{C}^{-1} (\mat{C}_{,i} \mat{C}^{-1} \mat{C}_{,j} + \mat{C}_{,j} \mat{C}^{-1} \mat{C}_{,i}) \mat{C}^{-1} \langle \mat{D} \rangle]. 
 \end{eqnarray}
The first term is the ordinary Fisher matrix, but we must replace $\mat{C}^{-1} \rightarrow \mat{C}_{\rm biased}^{-1}$. The second term contains $\mat{C}^{-1} \langle \mat{D} \rangle \rightarrow \mat{C}_{\rm biased}^{-1} (\mat{C} + \vect{x}_c \vect{x}_c^T)$. The width of the likelihood (as explored in MCMC) no longer has a clear interpretation because the model in the likelihood $\mat{C}_{\rm biased}$ does not coincide with $(\mat{C}_{\rm true} + \vect{x}_c \vect{x}_c^T)$. 

\bibliographystyle{apj}
\bibliography{robust_cmbpol}

\begin{thebibliography}{}
\expandafter\ifx\csname natexlab\endcsname\relax\def\natexlab#1{#1}\fi

\bibitem[{{Aiola} {et~al.}(2012){Aiola}, {Amico}, {Battaglia}, {Battistelli},
  {Ba{\'o}}, {de Bernardis}, {Bersanelli}, {Boscaleri}, {Cavaliere},
  {Coppolecchia}, {Cruciani}, {Cuttaia}, {D'Addabbo}, {D'Alessandro}, {De
  Gregori}, {Del Torto}, {De Petris}, {Fiorineschi}, {Franceschet},
  {Franceschi}, {Gervasi}, {Goldie}, {Gregorio}, {Haynes}, {Krachmalnicoff},
  {Lamagna}, {Maffei}, {Maino}, {Masi}, {Mennella}, {Morgante}, {Nati}, {Ng},
  {Pagano}, {Passerini}, {Peverini}, {Piacentini}, {Piccirillo}, {Pisano},
  {Ricciardi}, {Rissone}, {Romeo}, {Salatino}, {Sandri}, {Schillaci},
  {Stringhetti}, {Tartari}, {Tascone}, {Terenzi}, {Tomasi}, {Tommasi}, {Villa},
  {Virone}, {Withington}, {Zacchei}, \& {Zannoni}}]{2012SPIE.8446E..7AA}
{Aiola}, S., {Amico}, G., {Battaglia}, P., {et~al.} 2012, in \procspie, Vol.
  8446, Ground-based and Airborne Instrumentation for Astronomy IV, 84467A

\bibitem[{{Allison} {et~al.}(2015){Allison}, {Caucal}, {Calabrese}, {Dunkley},
  \& {Louis}}]{2015PhRvD..92l3535A}
{Allison}, R., {Caucal}, P., {Calabrese}, E., {Dunkley}, J., \& {Louis}, T.
  2015, \prd, 92, 123535

\bibitem[{{BICEP2 and Keck Array Collaborations} {et~al.}(2015){BICEP2 and Keck
  Array Collaborations}, {Ade}, {Ahmed}, {Aikin}, {Alexander}, {Barkats},
  {Benton}, {Bischoff}, {Bock}, {Brevik}, {Buder}, {Bullock}, {Buza},
  {Connors}, {Crill}, {Dowell}, {Dvorkin}, {Duband}, {Filippini}, {Fliescher},
  {Golwala}, {Halpern}, {Harrison}, {Hasselfield}, {Hildebrandt}, {Hilton},
  {Hristov}, {Hui}, {Irwin}, {Karkare}, {Kaufman}, {Keating}, {Kefeli},
  {Kernasovskiy}, {Kovac}, {Kuo}, {Leitch}, {Lueker}, {Mason}, {Megerian},
  {Netterfield}, {Nguyen}, {O'Brient}, {Ogburn}, {Orlando}, {Pryke},
  {Reintsema}, {Richter}, {Schwarz}, {Sheehy}, {Staniszewski}, {Sudiwala},
  {Teply}, {Thompson}, {Tolan}, {Turner}, {Vieregg}, {Weber}, {Willmert},
  {Wong}, \& {Yoon}}]{2015ApJ...811..126B}
{BICEP2 and Keck Array Collaborations}, {Ade}, P.~A.~R., {Ahmed}, Z., {et~al.}
  2015, \apj, 811, 126

\bibitem[{{Bond} {et~al.}(1998){Bond}, {Jaffe}, \&
  {Knox}}]{1998PhRvD..57.2117B}
{Bond}, J.~R., {Jaffe}, A.~H., \& {Knox}, L. 1998, \prd, 57, 2117

\bibitem[{{Bunn} {et~al.}(2003){Bunn}, {Zaldarriaga}, {Tegmark}, \& {de
  Oliveira-Costa}}]{2003PhRvD..67b3501B}
{Bunn}, E.~F., {Zaldarriaga}, M., {Tegmark}, M., \& {de Oliveira-Costa}, A.
  2003, \prd, 67, 023501

\bibitem[{Byrd {et~al.}(1995)Byrd, Lu, Nocedal, \& Zhu}]{byrd1995limited}
Byrd, R.~H., Lu, P., Nocedal, J., \& Zhu, C. 1995, SIAM Journal on Scientific
  Computing, 16, 1190

\bibitem[{{Das} {et~al.}(2014){Das}, {Louis}, {Nolta}, {Addison},
  {Battistelli}, {Bond}, {Calabrese}, {Crichton}, {Devlin}, {Dicker},
  {Dunkley}, {D{\"u}nner}, {Fowler}, {Gralla}, {Hajian}, {Halpern},
  {Hasselfield}, {Hilton}, {Hincks}, {Hlozek}, {Huffenberger}, {Hughes},
  {Irwin}, {Kosowsky}, {Lupton}, {Marriage}, {Marsden}, {Menanteau}, {Moodley},
  {Niemack}, {Page}, {Partridge}, {Reese}, {Schmitt}, {Sehgal}, {Sherwin},
  {Sievers}, {Spergel}, {Staggs}, {Swetz}, {Switzer}, {Thornton}, {Trac}, \&
  {Wollack}}]{2014JCAP...04..014D}
{Das}, S., {Louis}, T., {Nolta}, M.~R., {et~al.} 2014, JCAP, 4, 014

\bibitem[{{Efstathiou}(2004{\natexlab{a}})}]{2004MNRAS.348..885E}
{Efstathiou}, G. 2004{\natexlab{a}}, \mnras, 348, 885

\bibitem[{{Efstathiou}(2004{\natexlab{b}})}]{2004MNRAS.349..603E}
---. 2004{\natexlab{b}}, \mnras, 349, 603

\bibitem[{{Efstathiou}(2006)}]{2006MNRAS.370..343E}
---. 2006, \mnras, 370, 343

\bibitem[{{Errard} {et~al.}(2016){Errard}, {Feeney}, {Peiris}, \&
  {Jaffe}}]{2016JCAP...03..052E}
{Errard}, J., {Feeney}, S.~M., {Peiris}, H.~V., \& {Jaffe}, A.~H. 2016, \jcap,
  3, 052

\bibitem[{{Essinger-Hileman} {et~al.}(2014){Essinger-Hileman}, {Ali}, {Amiri},
  {Appel}, {Araujo}, {Bennett}, {Boone}, {Chan}, {Cho}, {Chuss}, {Colazo},
  {Crowe}, {Denis}, {D{\"u}nner}, {Eimer}, {Gothe}, {Halpern}, {Harrington},
  {Hilton}, {Hinshaw}, {Huang}, {Irwin}, {Jones}, {Karakla}, {Kogut}, {Larson},
  {Limon}, {Lowry}, {Marriage}, {Mehrle}, {Miller}, {Miller}, {Moseley},
  {Novak}, {Reintsema}, {Rostem}, {Stevenson}, {Towner}, {U-Yen}, {Wagner},
  {Watts}, {Wollack}, {Xu}, \& {Zeng}}]{2014SPIE.9153E..1IE}
{Essinger-Hileman}, T., {Ali}, A., {Amiri}, M., {et~al.} 2014, in \procspie,
  Vol. 9153, Millimeter, Submillimeter, and Far-Infrared Detectors and
  Instrumentation for Astronomy VII, 91531I

\bibitem[{{Farhang} {et~al.}(2013){Farhang}, {Bond}, {Dor{\'e}}, \&
  {Netterfield}}]{2013ApJ...771...12F}
{Farhang}, M., {Bond}, J.~R., {Dor{\'e}}, O., \& {Netterfield}, C.~B. 2013,
  \apj, 771, 12

\bibitem[{{Gjerl{\o}w} {et~al.}(2015){Gjerl{\o}w}, {Colombo}, {Eriksen},
  {G{\'o}rski}, {Gruppuso}, {Jewell}, {Plaszczynski}, \&
  {Wehus}}]{2015ApJS..221....5G}
{Gjerl{\o}w}, E., {Colombo}, L.~P.~L., {Eriksen}, H.~K., {et~al.} 2015, \apjs,
  221, 5

\bibitem[{{G{\'o}rski} {et~al.}(2005){G{\'o}rski}, {Hivon}, {Banday},
  {Wandelt}, {Hansen}, {Reinecke}, \& {Bartelmann}}]{2005ApJ...622..759G}
{G{\'o}rski}, K.~M., {Hivon}, E., {Banday}, A.~J., {et~al.} 2005, \apj, 622,
  759

\bibitem[{{Grain} {et~al.}(2009){Grain}, {Tristram}, \&
  {Stompor}}]{2009PhRvD..79l3515G}
{Grain}, J., {Tristram}, M., \& {Stompor}, R. 2009, \prd, 79, 123515

\bibitem[{{Groeneboom} {et~al.}(2009){Groeneboom}, {Eriksen}, {Gorski}, {Huey},
  {Jewell}, \& {Wandelt}}]{2009ApJ...702L..87G}
{Groeneboom}, N.~E., {Eriksen}, H.~K., {Gorski}, K., {et~al.} 2009, \apjl, 702,
  L87

\bibitem[{{Gruetjen} \& {Shellard}(2014)}]{2014PhRvD..89f3008G}
{Gruetjen}, H.~F., \& {Shellard}, E.~P.~S. 2014, \prd, 89, 063008

\bibitem[{{Gruppuso} {et~al.}(2009){Gruppuso}, {de Rosa}, {Cabella}, {Paci},
  {Finelli}, {Natoli}, {de Gasperis}, \& {Mandolesi}}]{2009MNRAS.400..463G}
{Gruppuso}, A., {de Rosa}, A., {Cabella}, P., {et~al.} 2009, \mnras, 400, 463

\bibitem[{{Hamimeche} \& {Lewis}(2009)}]{2009PhRvD..79h3012H}
{Hamimeche}, S., \& {Lewis}, A. 2009, \prd, 79, 083012

\bibitem[{{Henderson} {et~al.}(2016){Henderson}, {Allison}, {Austermann},
  {Baildon}, {Battaglia}, {Beall}, {Becker}, {De Bernardis}, {Bond},
  {Calabrese}, {Choi}, {Coughlin}, {Crowley}, {Datta}, {Devlin}, {Duff},
  {Dunkley}, {D{\"u}nner}, {van Engelen}, {Gallardo}, {Grace}, {Hasselfield},
  {Hills}, {Hilton}, {Hincks}, {Hlo{\^z}ek}, {Ho}, {Hubmayr}, {Huffenberger},
  {Hughes}, {Irwin}, {Koopman}, {Kosowsky}, {Li}, {McMahon}, {Munson}, {Nati},
  {Newburgh}, {Niemack}, {Niraula}, {Page}, {Pappas}, {Salatino}, {Schillaci},
  {Schmitt}, {Sehgal}, {Sherwin}, {Sievers}, {Simon}, {Spergel}, {Staggs},
  {Stevens}, {Thornton}, {Van Lanen}, {Vavagiakis}, {Ward}, \&
  {Wollack}}]{2016JLTP..184..772H}
{Henderson}, S.~W., {Allison}, R., {Austermann}, J., {et~al.} 2016, Journal of
  Low Temperature Physics, 184, 772

\bibitem[{{Hinshaw} {et~al.}(2009){Hinshaw}, {Weiland}, {Hill}, {Odegard},
  {Larson}, {Bennett}, {Dunkley}, {Gold}, {Greason}, {Jarosik}, {Komatsu},
  {Nolta}, {Page}, {Spergel}, {Wollack}, {Halpern}, {Kogut}, {Limon}, {Meyer},
  {Tucker}, \& {Wright}}]{2009ApJS..180..225H}
{Hinshaw}, G., {Weiland}, J.~L., {Hill}, R.~S., {et~al.} 2009, \apjs, 180, 225

\bibitem[{{Hinshaw} {et~al.}(2013){Hinshaw}, {Larson}, {Komatsu}, {Spergel},
  {Bennett}, {Dunkley}, {Nolta}, {Halpern}, {Hill}, {Odegard}, {Page}, {Smith},
  {Weiland}, {Gold}, {Jarosik}, {Kogut}, {Limon}, {Meyer}, {Tucker}, {Wollack},
  \& {Wright}}]{2013ApJS..208...19H}
{Hinshaw}, G., {Larson}, D., {Komatsu}, E., {et~al.} 2013, \apjs, 208, 19

\bibitem[{{Hivon} {et~al.}(2002){Hivon}, {G{\'o}rski}, {Netterfield}, {Crill},
  {Prunet}, \& {Hansen}}]{2002ApJ...567....2H}
{Hivon}, E., {G{\'o}rski}, K.~M., {Netterfield}, C.~B., {et~al.} 2002, \apj,
  567, 2

\bibitem[{{Hu} \& {White}(1997)}]{1997NewA....2..323H}
{Hu}, W., \& {White}, M. 1997, \na, 2, 323

\bibitem[{{Kamionkowski} {et~al.}(1997){Kamionkowski}, {Kosowsky}, \&
  {Stebbins}}]{1997PhRvD..55.7368K}
{Kamionkowski}, M., {Kosowsky}, A., \& {Stebbins}, A. 1997, \prd, 55, 7368

\bibitem[{{Kamionkowski} \& {Kovetz}(2014)}]{2014PhRvL.113s1303K}
{Kamionkowski}, M., \& {Kovetz}, E.~D. 2014, Physical Review Letters, 113,
  191303

\bibitem[{{Katayama} \& {Komatsu}(2011)}]{2011ApJ...737...78K}
{Katayama}, N., \& {Komatsu}, E. 2011, \apj, 737, 78

\bibitem[{{Kogut} {et~al.}(2011){Kogut}, {Fixsen}, {Chuss}, {Dotson}, {Dwek},
  {Halpern}, {Hinshaw}, {Meyer}, {Moseley}, {Seiffert}, {Spergel}, \&
  {Wollack}}]{2011JCAP...07..025K}
{Kogut}, A., {Fixsen}, D.~J., {Chuss}, D.~T., {et~al.} 2011, \jcap, 7, 025

\bibitem[{{Kouveliotou} {et~al.}(2014){Kouveliotou}, {Agol}, {Batalha}, {Bean},
  {Bentz}, {Cornish}, {Dressler}, {Figueroa-Feliciano}, {Gaudi}, {Guyon},
  {Hartmann}, {Kalirai}, {Niemack}, {Ozel}, {Reynolds}, {Roberge}, {Straughn},
  {Weinberg}, \& {Zmuidzinas}}]{2014arXiv1401.3741K}
{Kouveliotou}, C., {Agol}, E., {Batalha}, N., {et~al.} 2014, ArXiv e-prints,
  arXiv:1401.3741

\bibitem[{{Lazear} {et~al.}(2014){Lazear}, {Ade}, {Benford}, {Bennett},
  {Chuss}, {Dotson}, {Eimer}, {Fixsen}, {Halpern}, {Hilton}, {Hinderks},
  {Hinshaw}, {Irwin}, {Jhabvala}, {Johnson}, {Kogut}, {Lowe}, {McMahon},
  {Miller}, {Mirel}, {Moseley}, {Rodriguez}, {Sharp}, {Staguhn}, {Switzer},
  {Tucker}, {Weston}, \& {Wollack}}]{2014SPIE.9153E..1LL}
{Lazear}, J., {Ade}, P.~A.~R., {Benford}, D., {et~al.} 2014, in \procspie, Vol.
  9153, Millimeter, Submillimeter, and Far-Infrared Detectors and
  Instrumentation for Astronomy VII, 91531L

\bibitem[{{L{\'o}pez-Caniego} {et~al.}(2014){L{\'o}pez-Caniego}, {Rebolo},
  {Aguiar}, {G{\'e}nova-Santos}, {G{\'o}mez-Re{\~n}asco}, {Gutierrez},
  {Herreros}, {Hoyland}, {L{\'o}pez-Caraballo}, {Pelaez Santos}, {Poidevin},
  {Rubi{\~n}o-Mart{\'{\i}}n}, {Sanchez de la Rosa}, {Tramonte}, {Vega-Moreno},
  {Viera-Curbelo}, {Vignaga}, {Mart{\'{\i}}nez-Gonz{\'a}lez}, {Barreiro},
  {Casaponsa}, {Casas}, {Diego}, {Fern{\'a}ndez-Cobos}, {Herranz}, {Ortiz},
  {Vielva}, {Artal}, {Aja}, {Cagigas}, {Cano}, {de la Fuente}, {Mediavilla},
  {Ter{\'a}n}, {Villa}, {Piccirillo}, {Battye}, {Blackhurst}, {Brown},
  {Davies}, {Davis}, {Dickinson}, {Grainge}, {Harper}, {Maffei}, {McCulloch},
  {Melhuish}, {Pisano}, {Watson}, {Hobson}, {Lasenby}, {Saunders}, \&
  {Scott}}]{2014arXiv1401.4690L}
{L{\'o}pez-Caniego}, M., {Rebolo}, R., {Aguiar}, M., {et~al.} 2014, ArXiv
  e-prints, arXiv:1401.4690

\bibitem[{{Mangilli} {et~al.}(2015){Mangilli}, {Plaszczynski}, \&
  {Tristram}}]{2015MNRAS.453.3174M}
{Mangilli}, A., {Plaszczynski}, S., \& {Tristram}, M. 2015, \mnras, 453, 3174

\bibitem[{{Matsumura} {et~al.}(2014){Matsumura}, {Akiba}, {Borrill}, {Chinone},
  {Dobbs}, {Fuke}, {Ghribi}, {Hasegawa}, {Hattori}, {Hattori}, {Hazumi},
  {Holzapfel}, {Inoue}, {Ishidoshiro}, {Ishino}, {Ishitsuka}, {Karatsu},
  {Katayama}, {Kawano}, {Kibayashi}, {Kibe}, {Kimura}, {Kimura}, {Koga},
  {Kozu}, {Komatsu}, {Lee}, {Matsuhara}, {Mima}, {Mitsuda}, {Mizukami},
  {Morii}, {Morishima}, {Murayama}, {Nagai}, {Nagata}, {Nakamura}, {Naruse},
  {Natsume}, {Nishibori}, {Nishino}, {Noda}, {Noguchi}, {Ogawa}, {Oguri},
  {Ohta}, {Otani}, {Richards}, {Sakai}, {Sato}, {Sato}, {Sekimoto}, {Shimizu},
  {Shinozaki}, {Sugita}, {Suzuki}, {Suzuki}, {Tajima}, {Takada}, {Takakura},
  {Takei}, {Tomaru}, {Uzawa}, {Wada}, {Watanabe}, {Yoshida}, {Yamasaki},
  {Yoshida}, \& {Yotsumoto}}]{2014JLTP..176..733M}
{Matsumura}, T., {Akiba}, Y., {Borrill}, J., {et~al.} 2014, Journal of Low
  Temperature Physics, 176, 733

\bibitem[{{Page} {et~al.}(2007){Page}, {Hinshaw}, {Komatsu}, {Nolta},
  {Spergel}, {Bennett}, {Barnes}, {Bean}, {Dor{\'e}}, {Dunkley}, {Halpern},
  {Hill}, {Jarosik}, {Kogut}, {Limon}, {Meyer}, {Odegard}, {Peiris}, {Tucker},
  {Verde}, {Weiland}, {Wollack}, \& {Wright}}]{2007ApJS..170..335P}
{Page}, L., {Hinshaw}, G., {Komatsu}, E., {et~al.} 2007, \apjs, 170, 335

\bibitem[{{Peebles}(1973)}]{1973ApJ...185..413P}
{Peebles}, P.~J.~E. 1973, \apj, 185, 413

\bibitem[{{Planck Collaboration} {et~al.}(2014){Planck Collaboration}, {Ade},
  {Aghanim}, {Armitage-Caplan}, {Arnaud}, {Ashdown}, {Atrio-Barandela},
  {Aumont}, {Baccigalupi}, {Banday}, \& et~al.}]{2014A&A...571A..23P}
{Planck Collaboration}, {Ade}, P.~A.~R., {Aghanim}, N., {et~al.} 2014, \aap,
  571, A23

\bibitem[{{Planck Collaboration} {et~al.}(2015){Planck Collaboration},
  {Aghanim}, {Arnaud}, {Ashdown}, {Aumont}, {Baccigalupi}, {Banday},
  {Barreiro}, {Bartlett}, {Bartolo}, \& et~al.}]{2015arXiv150702704P}
{Planck Collaboration}, {Aghanim}, N., {Arnaud}, M., {et~al.} 2015, ArXiv
  e-prints, arXiv:1507.02704

\bibitem[{{Remazeilles} {et~al.}(2016){Remazeilles}, {Dickinson}, {Eriksen}, \&
  {Wehus}}]{2016MNRAS.458.2032R}
{Remazeilles}, M., {Dickinson}, C., {Eriksen}, H.~K.~K., \& {Wehus}, I.~K.
  2016, \mnras, 458, 2032

\bibitem[{{Rocha} {et~al.}(2011){Rocha}, {Contaldi}, {Bond}, \&
  {G{\'o}rski}}]{2011MNRAS.414..823R}
{Rocha}, G., {Contaldi}, C.~R., {Bond}, J.~R., \& {G{\'o}rski}, K.~M. 2011,
  \mnras, 414, 823

\bibitem[{{Rybicki} \& {Press}(1992)}]{1992ApJ...398..169R}
{Rybicki}, G.~B., \& {Press}, W.~H. 1992, \apj, 398, 169

\bibitem[{{Seljak}(1998)}]{1998ApJ...503..492S}
{Seljak}, U. 1998, \apj, 503, 492

\bibitem[{{Seljak} \& {Zaldarriaga}(1997)}]{1997PhRvL..78.2054S}
{Seljak}, U., \& {Zaldarriaga}, M. 1997, Physical Review Letters, 78, 2054

\bibitem[{{Story} {et~al.}(2013){Story}, {Reichardt}, {Hou}, {Keisler}, {Aird},
  {Benson}, {Bleem}, {Carlstrom}, {Chang}, {Cho}, {Crawford}, {Crites}, {de
  Haan}, {Dobbs}, {Dudley}, {Follin}, {George}, {Halverson}, {Holder},
  {Holzapfel}, {Hoover}, {Hrubes}, {Joy}, {Knox}, {Lee}, {Leitch}, {Lueker},
  {Luong-Van}, {McMahon}, {Mehl}, {Meyer}, {Millea}, {Mohr}, {Montroy},
  {Padin}, {Plagge}, {Pryke}, {Ruhl}, {Sayre}, {Schaffer}, {Shaw}, {Shirokoff},
  {Spieler}, {Staniszewski}, {Stark}, {van Engelen}, {Vanderlinde}, {Vieira},
  {Williamson}, \& {Zahn}}]{2013ApJ...779...86S}
{Story}, K.~T., {Reichardt}, C.~L., {Hou}, Z., {et~al.} 2013, \apj, 779, 86

\bibitem[{{Tajima} {et~al.}(2012){Tajima}, {Choi}, {Hazumi}, {Ishitsuka},
  {Kawai}, \& {Yoshida}}]{2012SPIE.8452E..1MT}
{Tajima}, O., {Choi}, J., {Hazumi}, M., {et~al.} 2012, in \procspie, Vol. 8452,
  Millimeter, Submillimeter, and Far-Infrared Detectors and Instrumentation for
  Astronomy VI, 84521M

\bibitem[{{Taylor} \& {Kitching}(2010)}]{2010MNRAS.408..865T}
{Taylor}, A.~N., \& {Kitching}, T.~D. 2010, \mnras, 408, 865

\bibitem[{{Tegmark}(1997)}]{1997PhRvD..55.5895T}
{Tegmark}, M. 1997, \prd, 55, 5895

\bibitem[{{Tegmark} \& {de Oliveira-Costa}(2001)}]{2001PhRvD..64f3001T}
{Tegmark}, M., \& {de Oliveira-Costa}, A. 2001, \prd, 64, 063001

\bibitem[{{Tegmark} {et~al.}(1998){Tegmark}, {Hamilton}, {Strauss}, {Vogeley},
  \& {Szalay}}]{1998ApJ...499..555T}
{Tegmark}, M., {Hamilton}, A.~J.~S., {Strauss}, M.~A., {Vogeley}, M.~S., \&
  {Szalay}, A.~S. 1998, \apj, 499, 555

\bibitem[{{Tegmark} {et~al.}(1997){Tegmark}, {Taylor}, \&
  {Heavens}}]{1997ApJ...480...22T}
{Tegmark}, M., {Taylor}, A.~N., \& {Heavens}, A.~F. 1997, \apj, 480, 22

\bibitem[{{Tristram} {et~al.}(2005){Tristram}, {Mac{\'{\i}}as-P{\'e}rez},
  {Renault}, \& {Santos}}]{2005MNRAS.358..833T}
{Tristram}, M., {Mac{\'{\i}}as-P{\'e}rez}, J.~F., {Renault}, C., \& {Santos},
  D. 2005, \mnras, 358, 833

\bibitem[{{Watts} {et~al.}(2015){Watts}, {Larson}, {Marriage}, {Abitbol},
  {Appel}, {Bennett}, {Chuss}, {Eimer}, {Essinger-Hileman}, {Miller}, {Rostem},
  \& {Wollack}}]{2015ApJ...814..103W}
{Watts}, D.~J., {Larson}, D., {Marriage}, T.~A., {et~al.} 2015, \apj, 814, 103

\bibitem[{Zhu {et~al.}(1997)Zhu, Byrd, Lu, \& Nocedal}]{zhu1997algorithm}
Zhu, C., Byrd, R.~H., Lu, P., \& Nocedal, J. 1997, ACM Transactions on
  Mathematical Software (TOMS), 23, 550

\end{thebibliography}

\end{document}